\documentclass[twocolumn]{aastex631}

\usepackage{amsmath,amsfonts,amssymb}
\usepackage{graphicx}

\submitjournal{ApJ}
\accepted{January 29, 2024}

\shorttitle{Imaging with Photonic Lanterns}
\shortauthors{Kim et al.}

\graphicspath{{./}{figures/}}

\begin{document}

\title{Coherent Imaging with Photonic Lanterns}

\correspondingauthor{Yoo Jung Kim}
\email{yjkim@astro.ucla.edu}

\author[0000-0003-1392-0845]{Yoo Jung Kim}
\affiliation{Department of Physics \& Astronomy, 430 Portola Plaza, University of California, Los Angeles, CA 90095, USA}
\author[0000-0002-0176-8973]{Michael P. Fitzgerald}
\affiliation{Department of Physics \& Astronomy, 430 Portola Plaza, University of California, Los Angeles, CA 90095, USA}
\author[0000-0001-8542-3317]{Jonathan Lin}
\affiliation{Department of Physics \& Astronomy, 430 Portola Plaza, University of California, Los Angeles, CA 90095, USA}
\author[0000-0001-6871-6775]{Steph Sallum}
\affiliation{Department of Physics \& Astronomy, University of California Irvine, 4129 Frederick Reines Hall, Irvine, CA 92697, USA}
\author[0000-0002-6171-9081]{Yinzi Xin}
\affiliation{Department of Astronomy, California Institute of Technology, Pasadena, CA 91125, USA}
\author[0000-0001-5213-6207]{Nemanja Jovanovic}
\affiliation{Department of Astronomy, California Institute of Technology, Pasadena, CA 91125, USA}
\author[0000-0002-5606-3874]{Sergio Leon-Saval}
\affiliation{Sydney Astrophotonic Instrumentation Laboratory, School of Physics, The University of Sydney, Sydney, NSW 2006, Australia}

\begin{abstract}

Photonic Lanterns (PLs) are tapered waveguides that gradually transition from a multi-mode fiber geometry to a bundle of single-mode fibers (SMFs). They can efficiently couple multi-mode telescope light into a multi-mode fiber entrance at the focal plane and convert it into multiple single-mode beams. Thus, each SMF samples its unique mode (lantern principal mode) of the telescope light in the pupil, analogous to subapertures in aperture masking interferometry (AMI). Coherent imaging with PLs can be enabled by interfering SMF outputs and applying phase modulation, which can be achieved using a photonic chip beam combiner at the backend (e.g., the ABCD beam combiner). In this study, we investigate the potential of coherent imaging by interfering SMF outputs of a PL with a single telescope. We demonstrate that the visibilities that can be measured from a PL are mutual intensities incident on the pupil weighted by the cross-correlation of a pair of lantern modes. From numerically simulated lantern principal modes of a 6-port PL, we find that interferometric observables using a PL behave similarly to separated-aperture visibilities for simple models on small angular scales ($<\lambda/D$) but with greater sensitivity to symmetries and capability to break phase angle degeneracies. Furthermore, we present simulated observations with wavefront errors and compare them to AMI. Despite the redundancy caused by extended lantern principal modes, spatial filtering offers stability to wavefront errors.
Our simulated observations suggest that PLs may offer significant benefits in the photon noise-limited regime and in resolving small angular scales at low contrast regime.

\end{abstract}

\keywords{Direct imaging (387) --- Optical interferometry (1168) --- High angular resolution (2167) --- Astronomical techniques (1684) --- Circumstellar disks (235)}

\section{Introduction}\label{sec:intro}

\begin{figure*}
    \centering
    \includegraphics[scale=0.65]{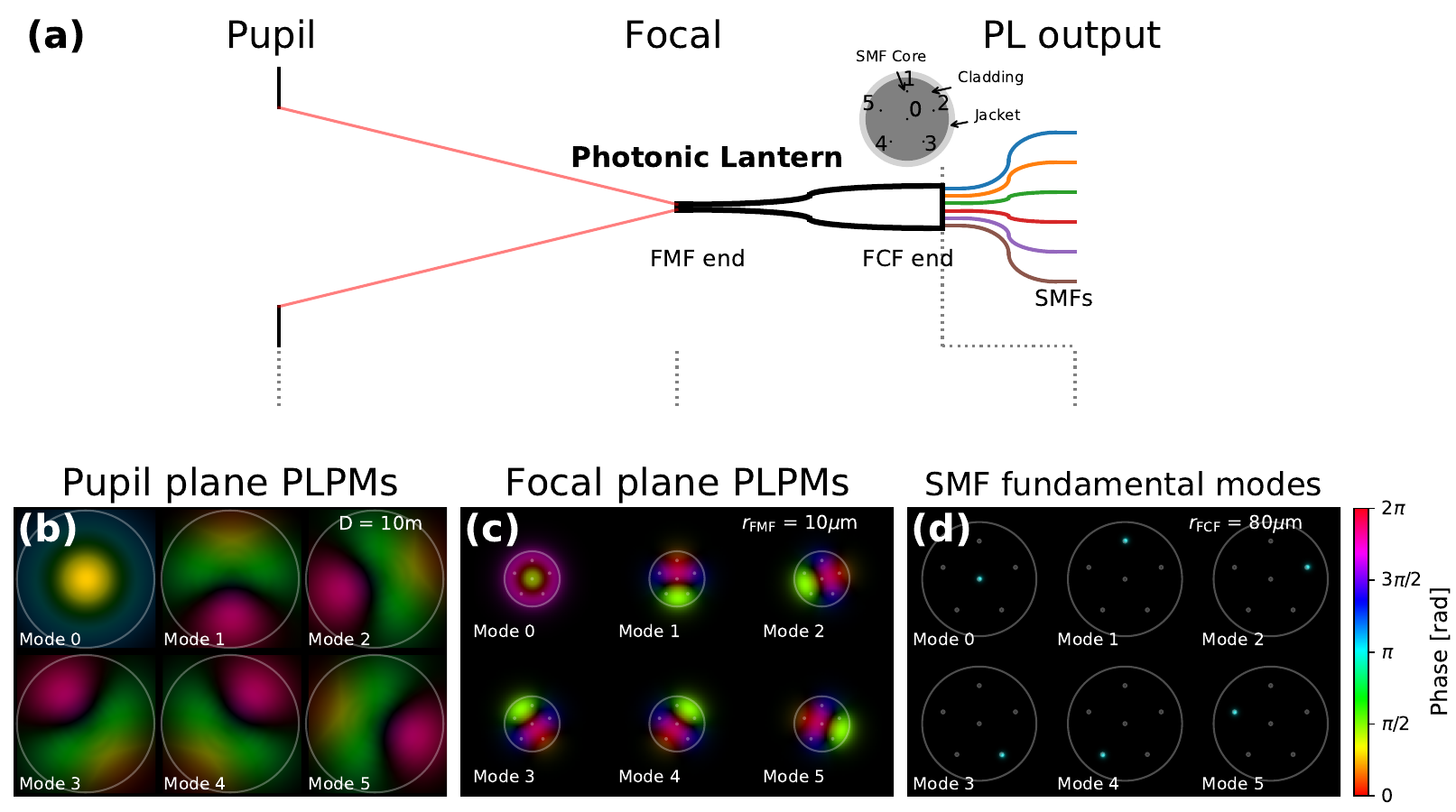}
    \caption{(a) A schematic diagram of a standard 6-port PL. The telescope light is injected into the PL at the focal plane through the FMF end, propagates through the PL and becomes confined in the SMFs. The SMF outputs can be directed to photonic chips or other devices and then to the detector. A cross-section of the PL at the FCF end is shown in the upper right. The simulated structures of six supported PL principal modes (modes 0 through 5) are shown in the lower panels, evaluated at the (b) pupil plane, (c) focal plane (FMF end), and (d) FCF end. Colors represent the phase and saturation indicates the amplitude. The white circles in (b) show the size of the telescope aperture where we assume a circular geometry with a diameter of 10\,m. The white circles in (c) and (d) show the geometry of the PL, with small circles indicating the size of the SMF core and large circles denoting the size of the inner jacket.
    See Section \ref{ssec:simulation} for more details of the simulation setup.   
    }
    \label{fig:PL}
\end{figure*}

Astronomical adaptive optics (AO) enables  diffraction-limited resolution ($\sim\lambda/D$) in the near infrared. However, in high-contrast applications, detecting a companion or resolving structures at such separations remains challenging, due to residual aberrations in AO-corrected wavefronts. Indeed, searching for directly imageable planets is typically limited to separations larger than a few $\lambda/D$ \citep{guy13, rua17}. Decreasing the AO-corrected inner working angle via novel imaging techniques will enable detections of extrasolar planets on tighter orbits around more distant stars.

There have been efforts to push to the ultimate limits set by diffraction using interferometric methods with a single telescope, which can reach resolutions as small as $\lambda/2D$. Examples include techniques such as non-redundant masking (NRM) interferometry \citep{bal86, han87, tut00, sal19} and filled-aperture kernel phase interferometry \citep{mar10}. NRM uses a pupil mask with a several machined holes designed so that the baselines formed by each pair correspond to unique spatial frequencies. The images formed at the focal plane are interferometric fringes, and complex visibilities can be extracted from the Fourier-transformed images. However, the mask blocks most of the light at the pupil, trading visibility measurement fidelity with photon sensitivity. On the other hand, kernel phase interferometry does not block the light, but rather treats the telescope pupil as an array of sub-apertures that form redundant baselines. While this technique is efficient in terms of random photon noise, it is effective only in the high-Strehl regime, where instrumental pupil-plane phases are small enough to linearly perturb the focal plane image \citep{ire13}. The pupil remapping technique (fibered aperture masking) is another approach, which divides the telescope pupil into multiple subapertures each feeding a single-mode fiber (SMF) \citep{per06, hub12, jov12}. In this way the entire telescope pupil can be used, achieving better Fourier coverage than the NRM. However, it still requires a high Strehl ratio to ensure modest coupling into SMFs, and a complex beam combination structure is required to measure interferometric properties from many baselines. 

In this work, we study the potential of achieving the resolution comparable to interferometry using a photonic device called the photonic lantern (PL) on a single telescope. PLs are tapered waveguides that gradually transition from a few-mode fiber (FMF) geometry to a bundle of SMFs, or a few-core fiber (FCF) \citep{leo13,bir15}. Figure \ref{fig:PL}(a) shows a schematic of a PL. When the AO-corrected telescope light couples into the FMF end of the PL at the focal plane, it becomes confined within the cores as it propagates through the lantern transition. Therefore, a PL splits a few-moded wavefront into a few single-moded beams, each in an individual SMF. The outputs confined in the SMFs are highly stable due to the spatial filtering nature of the SMF \citep{jov16}, making it ideal for observations that require high fidelity. The output SMFs can be used as an input for multiple devices, for example for a high-resolution spectrometer \citep{jov16, lin21b}, a focal-plane wavefront sensor \citep{nor20, lin22a, lin23b}, or for other photonic devices. If the PL were mode-selective \citep{leo14}, it could be used as a nuller that suppresses the on-axis starlight while coupling the off-axis light from a companion source \citep{xin22}.

Here, we consider interfering the SMF outputs to exploit the coherence properties of a PL. This can be realized by feeding SMF outputs into a backend photonic integrated circuit beam combiner. We illustrate this concept in more detail and describe interferometric observables that can be measured by PLs in \S\ref{sec:concept}. 
In \S\ref{sec:WFE}, we investigate the effects of WFEs on measurement of PL interferometry observables, specifically closure phases.
In \S\ref{sec:potential}, we perform mock observations and demonstrate the potential of PL interferometry for parametric modeling (visibility fitting) and for non-parametric image reconstruction. Finally, in \S\ref{sec:discussion}, we discuss prospects of this technique and future exploration directions.

\section{Visibilities measured by Photonic Lanterns}\label{sec:concept}

In classical interferometry, pairs of apertures are interfered to measure the corresponding fringe visibility. For PLs, the lantern principal modes, which we will define in \S\ref{ssec:concept}, are analogous to individual apertures in classical interferometry. 
We develop the corresponding PL visibility observables in \S \ref{ssec:plvis} and simulate these observables for simple system geometries in \S \ref{ssec:simulation}.

\subsection{Lantern Principal Modes as Effective Apertures}\label{ssec:concept}

For a radially symmetric weakly guiding step-index waveguide, the linearly polarized modes (LP modes; ${\rm LP_{lm(a,b)}}$) describe the propagation eigenmodes. Ignoring the dual multiplicity introduced by polarization, an $N$-moded fiber supports $N$ modes of propagation and their spatial modes can be described as superposition of $N$ LP modes. When telescope light is injected into an SMF at the focal plane, a portion of the light matching the mode structure of the SMF ($\rm{LP_{01}}$; Gaussian-like intensity profile and flat phase front) will couple and propagate through the fiber. Likewise, when telescope light is injected into an $N$-moded PL entrance (FMF end) in the focal plane, a fraction of the light that matches the first $N$ LP modes gets coupled and propagates through the PL to the SMF end. Therefore the $N$ modes in the $N$-moded PL entrance are mapped to $N$ SMF outputs. 

For PLs, we can define {\it Photonic Lantern Principal Modes} (PLPMs; ``lantern modes'' in \citet{lin21b}) --- a different orthogonal superposition of the spatial ${\rm LP_{lm(a,b)}}$ excited modes that maps to one SMF output. Therefore PLPMs can be determined by back-propagating fundamental modes in the single-mode outputs to the PL entrance. The number of principal modes is equal to the number of supported LP modes (${\rm LP_{lm(a,b)}}$) and lantern single-mode outputs. Principal mode propagation in a multimode fiber is a localization of a group of guided modes at the output in both the frequency and time domains \citep{nol21}. Based on the physical properties of the photonic lantern mode converter, the principal modes of the lantern do not suffer from spatio-temporal scattering along the short lantern transition and form orthogonal bases at both the input and the output ends of the waveguide. It is important to note that this definition is only an analogy to principal modes in multimode optical fiber transmission \citep{fan05}, in which principal modes are independent of wavelength and represent Eisenbud-Wigner-Smith states of the multimode transmission system \citep{car15}. In our case, they are wavelength dependent due to the optical transmission matrix solutions. However due to the nature of the photonic lantern transitions their temporal dispersion is negligible (wavelength independent) and can be considered constant.

Examples of the PLPMs for a standard 6-port lantern are shown in Figure \ref{fig:PL} (c). The fundamental modes confined in the SMFs (panel (d)), which form a modal basis in the PL output end, are back-propagated numerically to the lantern entrance.  Details of the simulation are presented in section \ref{ssec:simulation}.
The coupling efficiency of the complex focal-plane electric field into each of the PLPMs determines the flux of each SMF output.

The ``pupil plane PLPMs'', shown in panel (b), are defined as focal plane (PL entrance) PLPMs back-propagated to the pupil plane. The complex conjugate of these can be interpreted as effective apertures seen by the lantern. For instance, the pupil-plane wavefront that matches the pupil plane PLPM $i$ will couple into the PL and end up in the $i$-th SMF in the output, encoded as the complex amplitude of the fundamental mode. Therefore, although a single telescope aperture is used for collecting the light, SMF outputs of a PL represent the telescope light filtered by unique and spatially orthogonal effective apertures.

Examining these PLPMs in the pupil and focal planes, in general each mode has its unique phase and amplitude structure. Mode 0 couples an on-axis symmetric field while other modes are more sensitive to asymmetries. Also, each mode amplitude distribution peaks in different locations at the pupil, implying that each SMF preferentially samples more light at specific locations, similar to a ``subaperture'' in aperture masking interferometry (AMI). The complex-valued nature of the aperture function is a departure from traditional aperture masks (amplitude masks), although recent techniques like holographic aperture masks \citep{doe18, doe21} allow for complex-valued pupil functions. The interpretation of PL principal modes as effective apertures leads us to the concept of PL interferometry: interfering SMF outputs to learn about coherence properties of the source field distribution. 

\subsection{PL Visibilities}\label{ssec:plvis}

In conventional interferometry, the measurement process is aimed at sampling the complex visibilities of a source field distribution in the $u-v$ plane from light transmitted by two well-separated apertures that is then interfered. If the two apertures are placed at ($x_1$, $y_1$) and ($x_2$, $y_2$), 
the baseline is defined as ($B_u$, $B_v$) = ($x_2 - x_1$, $y_2 - y_1$). By interfering the fields at the two spatial points, the visibility at the spatial frequency $(u, v) = (B_u / \lambda, B_v / \lambda$) can be sampled and is expressed as 
a normalized mutual intensity as follows:
\begin{equation}\label{eq:Jp}
    \mathcal{V}(u, v) \propto J_p(u, v) = \langle E_p(x_1, y_1; t) E_p^*(x_2, y_2; t)\rangle_t.
\end{equation}
\noindent 
$J_p(u,v)$ is the mutual intensity of the source at the two apertures, which is a measure of the correlation between the fields at different points in space. The mutual intensity is the mutual coherence function under the quasi-monochromatic assumption, ignoring the effects of temporal coherence.
$E_p(x,y)$ represents the source field at the pupil plane. The brackets represent time average. Then the source intensity distribution corresponds to a two-dimensional Fourier transform of $J_p(u, v)$ by the Van Cittert-Zernike theorem. By sampling a number of $\mathcal{V}(u, v)$ values in the $u-v$ plane (filling the $u-v$ plane), the source intensity distribution can be recovered.

However, for PLs, the effective apertures are not well-separated (Figure \ref{fig:PL}(b)). They spatially overlap each other and have unique complex structures. 
Therefore, defining a baseline is ambiguous and $J_p(u, v)$ cannot be directly sampled as in separated-aperture interferometry. 
In the following subsections, we examine what is observable by interfering two SMF outputs of a PL and how it relates to the source field and its mutual intensity function.

\subsubsection{Development of PL Interferometry Observables}\label{ssec:definition}

We have discussed in the previous subsection that pupil plane PLPMs represent the effective apertures of each SMF output. The effective pupil function of the $i$-th SMF output can be calculated by the telescope aperture function multiplied by the complex conjugate of the $i$-th pupil plane PLPM.
Then the complex amplitudes of the scalar electric fields in the SMF outputs can be described as an overlap integral in the pupil plane ($x$, $y$):
\begin{align}\label{eq:field}
    E_i(t) = \iint_{-\infty}^{\infty} E_{\rm p}(x,y; t) P_{{\rm eff}, i}(x, y) dx dy
\end{align}
where $E_{\rm p}(x,y; t)$ is the source field at the pupil plane and $P_{\rm eff, i}(x,y)$ is the effective pupil function of the $i$-th output. The coupling efficiency of the $i$-th SMF is then described as the normalized form of the absolute square of the above.

Now consider interfering light from a pair of outputs. Writing the electric fields in the two outputs as $E_1(t)$ and $E_2(t)$, intensities in each output SMF port can be written as 
\begin{align}
\begin{aligned}
    I_1 &= \langle E_1(t) E_1^*(t)\rangle_t \\
    I_2 &= \langle E_2(t) E_2^*(t)\rangle_t
    \end{aligned}
\end{align}
If we, for example, sum the two electric fields (coherently combine the two SMF outputs), the resulting intensity is then
\begin{align}
\begin{aligned}
\langle |E_1(t) + E_2(t)|^2\rangle_t &= I_1 + I_2 + 2{\rm Re}(\langle E_1(t)E_2^*(t)\rangle_t) \\
&= I_1 + I_2 + 2{\rm Re}(J_{12}).
\end{aligned}
\end{align}
The last term is the mutual intensity term, which is the modulating term depending on the coherence properties of $E_1$ and $E_2$. This term is related to the complex fringe visibility between the two SMF outputs as follows:
\begin{equation}\label{eq:visibility}
    \mathcal{V}_{12} = \frac{2}{I_1 + I_2} J_{12}.
\end{equation}
This is the visibility that can be measured by interfering SMF outputs of a PL, which we define as the {\it PL visibility}. For a 6-port PL, there are 15 unique pairs of ports, so 15 such complex visibilities can be defined ($\mathcal{V}_{ij}$). In the example of coherently combining the two SMF outputs, the real part of the complex visibility is determined.

In practice, both the real and the complex parts of the complex visibility $\mathcal{V}_{ij}$ can be determined using an ABCD beam combiner, which gets two inputs and provides four outputs with phase difference of multiples of 90 degrees \citep{sha77, ben09}. 
The four outputs from one ABCD beam combiner with inputs $E_i$ and $E_j$ would be 
\begin{align}
    \begin{aligned}
    I_{ij, A} &= |E_i + E_j|^2 = I_i + I_j + 2{\rm Re} (J_{ij}) \\ 
    I_{ij, B} &= |E_i + E_j e^{i\pi}|^2 = I_i + I_j + 2{\rm Re}(J_{ij} e^{-i\pi}) \\
    I_{ij, C} &= |E_i + E_j e^{i\pi/2}|^2 = I_i + I_j + 2{\rm Re}(J_{ij} e^{-i\pi/2}) \\
    I_{ij, D} &= |E_i + E_j e^{i3\pi/2}|^2 = I_i + I_j + 2{\rm Re}(J_{ij} e^{-i3\pi/2}).
    \end{aligned}\label{eq:ABCD}
\end{align}
Therefore, it is possible to solve for the complex visibility $\mathcal{V}_{ij}$ given the four outputs. In the above case, 
\begin{equation}\label{eq:ABCD2vis}
    \mathcal{V}_{ij} = \frac{(I_{ij,A} - I_{ij,B}) + (I_{ij,C} - I_{ij,D})i}{I_{ij,A} + I_{ij,B} + I_{ij,C} + I_{ij,D}}.
\end{equation}
Note that this assumes an ideal case where instrumental phase shift, instrumental contrast, and intensity losses are negligible, and equal splitting ratios are assumed. These should be carefully calibrated in order to achieve a high sensitivity \citep{col99, ben09}. Additionally, we assume a monochromatic, scalar electric field and defer the consideration of polarization and chromaticity to future work.

For a 6-port PL, light in each SMF output is split five ways and sent to ABCD beam combiners for pairwise beam combination. Therefore 15 ABCD beam combiners are needed, leading to 60 outputs in total. The outputs may additionally be split based on polarization. The outputs can be directed to an optional dispersing spectrometer and then to a detector.

\subsubsection{Spatial frequencies sampled by PL visibilities}

\begin{figure*}[hbt!]
    \centering
    \includegraphics[width=1\linewidth]{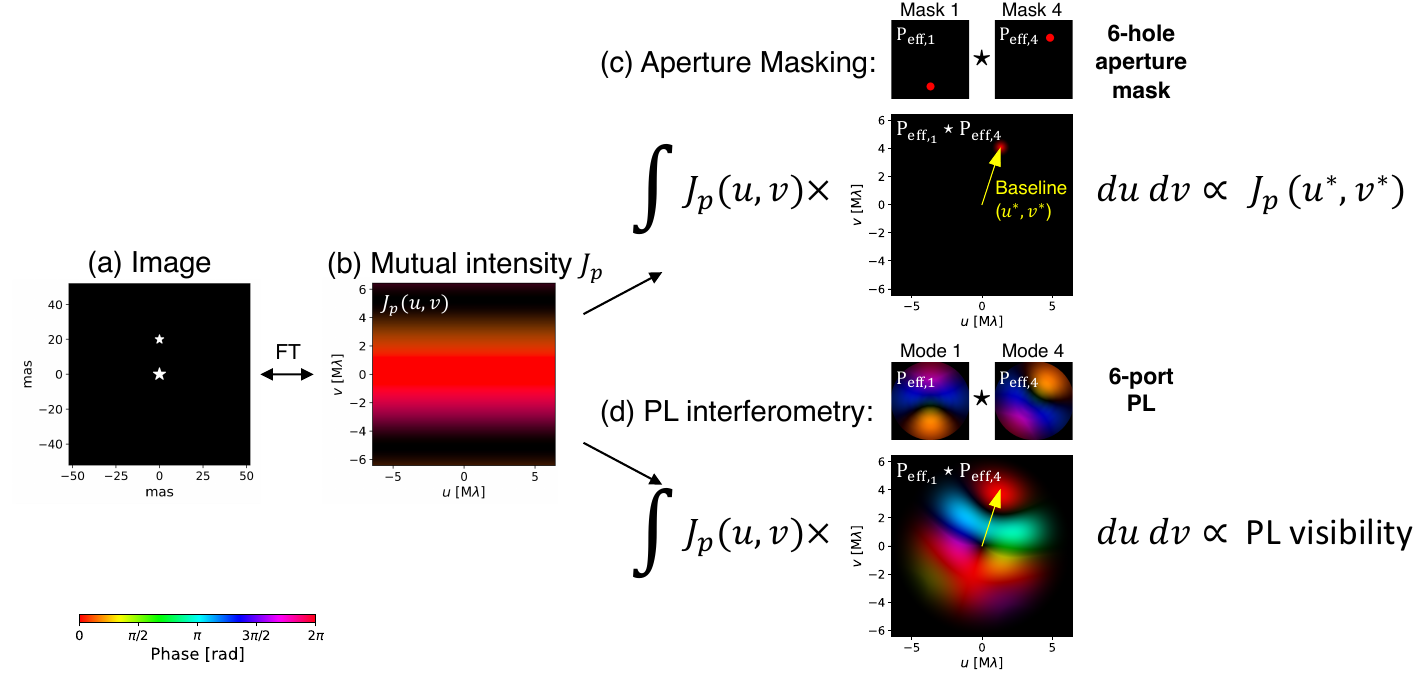}
    \caption{A concept diagram of PL interferometry, showing physical interpretation of PL visibilities. (a) An example scene consisting of two point sources separated by 20 mas and with flux ratio of 0.5, indicated by star symbols. (b) Mutual intensity in the pupil plane $J_p (u, v)$, which is the two-dimensional Fourier transform of the image in the left. (c) Cross-correlation of two subapertures of a 6-hole aperture mask. The hole locations are defined as the peaks in the PL pupil functions (see Figure \ref{fig:ccpupil}). The cross-correlated aperture functions are well-localized at some ($u^*$,$v^*$), so the visibility measured from light transmitted by these two apertures approximately equals to the visibility at this ($u^*$,$v^*$) baseline. (d) Cross-correlation of two PL pupil functions of a 6-port PL. The cross-correlated pupil functions is extended but has a defined blob at the ($u^*$,$v^*$) location in the aperture mask case. Visibilities we can measure from a PL correspond to the product of $J_p(u,v)$ and the cross-correlated pupil function integrated over the $u-v$ plane. The panels (b) and (d) are complex-valued: colors represent the phase and saturation indicates the amplitude in plots.}
    \label{fig:plvis}
\end{figure*}

\begin{figure*}[hbt!]
    \centering
    \includegraphics[width=0.9\linewidth]{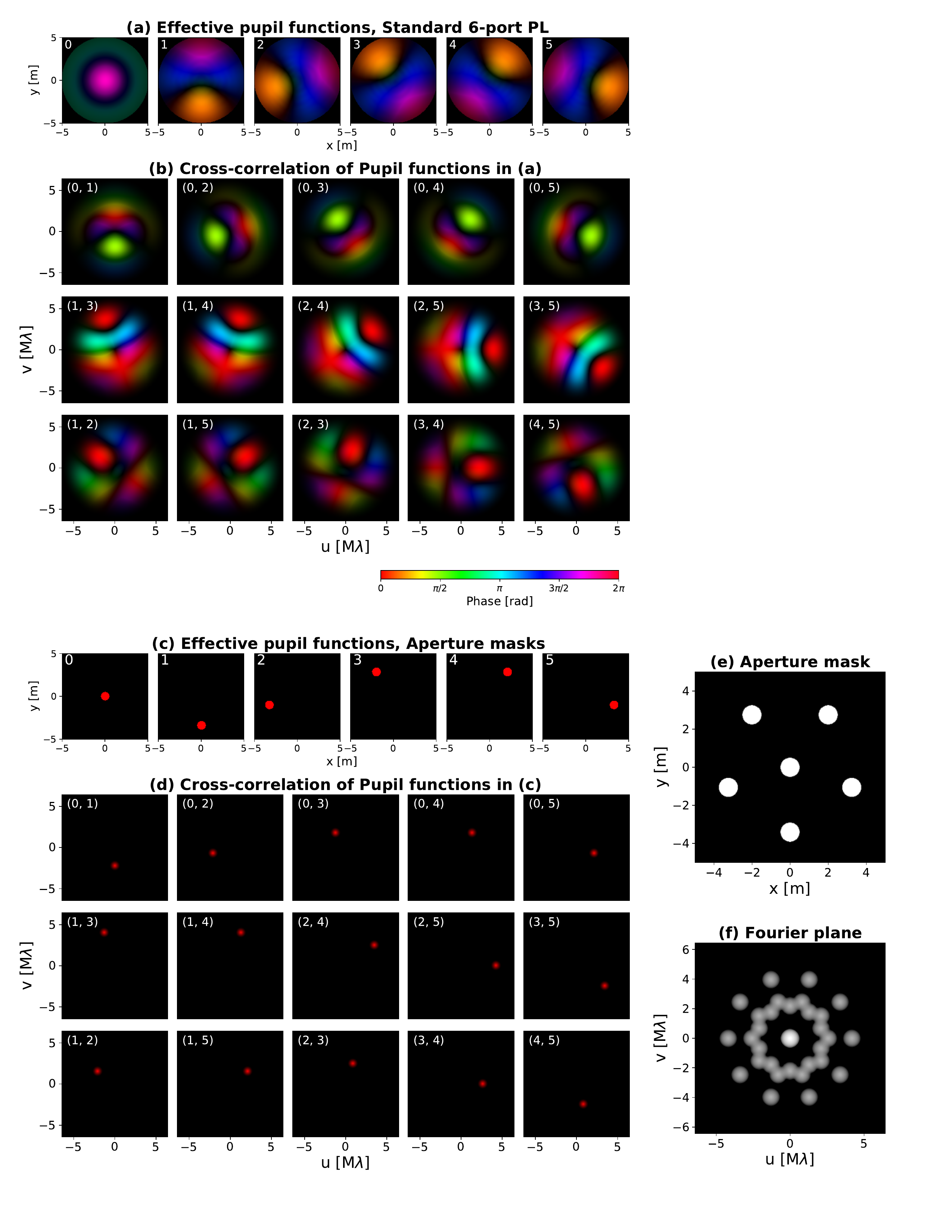}
    \caption{
    (a) Effective pupil functions of the supported six modes of a standard 6-port PL. The extent of each image is 10m $\times$ 10m. (b) Cross-correlation maps of the pupil functions in (a), where indices are indicated on the upper left. The cross-correlation of a pair of apertures can be interpreted as OTF of a pair of apertures (OTF$_{ij}$), which describes the spatial frequency that is sampled by interfering the two spatially filtered apertures. The wavelength $\lambda$ used  is 1.55$\mu m$. 
    (c) Effective pupil functions of the six subapertures in AMI, where the locations are matched to the location where effective pupil functions in (a) have maximum amplitudes. The extent of each image is 10m $\times$ 10m and each subaperture has a diameter of 1m.
    (d) Cross-correlation maps of the pupil functions in (c).
    Colors represent the phase and saturation indicates the amplitude.
    (e) The aperture mask configuration and (f) corresponding Fourier-plane in log scale (Fourier transform of the focal plane point spread function).
    }
    \label{fig:ccpupil}
\end{figure*}

In this subsection, we examine the spatial frequency information contained in the PL visibilities: relating PL visibilities to the mutual intensity at the pupil plane, $J_p(u, v)$.

Using Equation \ref{eq:field}, the relation between the mutual intensity measured from a pair of outputs ($J_{12}$) and the source mutual intensity can be established as follows:
\begin{align}\label{eq:J12}
    \begin{aligned}
        J_{ij} &= \langle E_i(t) E_j^*(t)\rangle_t \\
               &= 
               \iiiint_{-\infty}^{\infty} 
               \langle E_{\rm p}(x_i, y_i) E_{\rm p}^*(x_j, y_j)\rangle_t \\
               &\;\;\;\;\;\times P_{\rm eff,i}(x_i, y_i)  
               P_{\rm eff,j}^*(x_j, y_j) dx_i dy_i dx_j dy_j \\
               &= \iint_{-\infty}^{\infty} J_{\rm p}(u, v)~ (P_{\rm eff,j} \star P_{\rm eff,i})(u,v) du dv
    \end{aligned}
\end{align}
where $\star$ denotes the two-dimensional cross correlation. 
Therefore, the PL visibility between two SMF outputs $i$ and $j$ is then 
\begin{equation}\label{eq:vis_full}
    \mathcal{V}_{ij} = \frac{2 \iint_{\infty}^{\infty} J_{\rm p}(u, v) \; (P_{{\rm eff},j} \star P_{{\rm eff},i}) du dv}{\iint_{\infty}^{\infty} J_{\rm p}(u, v) \; (P_{{\rm eff},i} \star P_{{\rm eff},i} + P_{{\rm eff},j} \star P_{\rm eff,j}) du dv}
\end{equation}
Equation \ref{eq:vis_full} shows that the PL visibilities are proportional to the product of the mutual intensity in the pupil plane and the {\it cross-correlation of the effective pupil functions}, integrated over the $u-v$ plane.

 Note that the cross-correlation of pupil functions have correspondence to the optical transfer function (OTF). The OTF describes how an imaging system samples the spatial frequency ($u$, $v$) of the object intensity. If considering an image formed by a single aperture, the OTF is defined as the normalized autocorrelation of the pupil function: the image consists of sum of all the interferometric fringes formed by every pair of points within the aperture. 
On the other hand, if spatially filtered wavefronts from two subapertures are interfered, there is no interference between two points within a subaperture. Instead of combining the two pupil functions and taking its autocorrelation, we only need to take the cross-correlation of the two pupil functions to define an OTF of a pair of subapertures:
\begin{equation}\label{eq:cc}
    {\rm OTF}_{i,j} (u, v) \propto (P_{\rm eff,j} \star P_{\rm eff, i})(u, v).
\end{equation}
It is important to note that spatial filtering means that measurable mutual intensity $J_{ij} = \langle E_i E_j^* \rangle_t$ is a spatial frequency averaged value.

Figure \ref{fig:plvis} visualizes the interpretation of PL visibilities. 
Panel (a) shows an example scene of binary stars, separated by 20 mas. Panel (b) represents the mutual intensity in the pupil plane ($J_p(u,v)$), which is the two-dimensional Fourier transform of the image. In panel (d) we show PL pupil functions of mode 1 and 4 as an example, and their cross-correlation (OTF).  
The visibility measurable by interfering SMF outputs 1 and 4 corresponds to the mutual intensity distribution $J_p$ weighted by this cross-correlated pupil functions, normalized by the average intensity of outputs 1 and 4.
The cross-correlated pupil function has a complex structure, with a flat phase blob at some ($u^*$,$v^*$) location as indicated by the yellow arrow.

\subsubsection{Comparison to Aperture Masking Interferometry}\label{ssec:comparison}

In panel (c) of Figure \ref{fig:plvis}, we display an AMI analog case --- the case if the pupil functions were circular apertures of 1m diameter, placed at the location where PL pupil mode amplitudes are at maximum. The cross-correlation of pupil functions is well-localized in the $u-v$ plane because the subapertures are small compared to the baseline length. Thus, weighting by the cross-correlated pupil functions is essentially sampling the mutual intensity at the corresponding baseline (Equation \eqref{eq:Jp}). All the effective pupil functions and cross-correlated pupil functions for our simulated PL and the aperture mask are shown in Figure \ref{fig:ccpupil}.

Note the coincidence of the location of the flat phase blobs in the cross-correlated pupil functions of the PL and of the AMI analog case. Despite the complex structure of the PL cross-correlated pupil functions, the structure other than the blob is somewhat symmetric such that PL visibilities are dominated by the mutual intensity at the location of the blobs. From this we can infer that AMI is a reasonable analog for PL interferometry. This is valid if the mutual intensity at the pupil plane changes gradually over the $u-v$ plane (or small angular size object in the image plane), slower than the variation of the cross-correlated pupil functions over the $u-v$ plane. This is in line with the fact that the field of view of the PL is limited by the size of the FMF entrance at the focal plane: any off-axis light that does not fall onto the FMF entrance will not couple into the PL. Note that the limited field of view also implies that the PL pupil functions are extended. In AMI, the size of the subapertures determines the field of view.

Although PLs have limited field of view, there are a few potential advantages over AMI. First, the throughput is higher by several factors compared to conventional AMI with adaptive optics, because the aperture is not blocked by masks. Also, the relative intensities in the output SMFs, which are sensitive to low-order aberrations \citep{lin22a, kim22}, may be used to constrain source intensity distribution in addition to visibilities.

\subsection{Simulated Interferometric Observables with PLs}\label{ssec:simulation}

To assess the information content of PL visibilities, we numerically calculate PL visibilities for simple cases and compare with those of the AMI analog. We also investigate the stability of PL visibilities against wavefront errors (WFEs) in \S\ref{sec:WFE} and demonstrate their potential for parametric modeling and image reconstruction in \S\ref{sec:potential}.

\subsubsection{Simulation setup}

The simulated lantern is a standard 6 port lantern of which all SMFs have the same core refractive index and radius. We use a cladding index of 1.444, cladding-jacket index contrast of $5.5\times 10^{-3}$, and core-cladding index contrast of $8.8\times10^{-3}$. Each SMF core diameter is chosen to be 4.4~\textmu m and the FMF entrance diameter to be 10~\textmu m. The lantern taper length is set to 2~cm and the taper scale (scale difference between the output and input ends) is set to 8. See Figure \ref{fig:PL} for the geometry of the PL.

To determine the mode structure in the focal plane and the pupil plane (Figure \ref{fig:PL}(b) and (c)), we perform numerical beam propagation of fundamental modes in the SMFs. The simulations are monochromatic, with wavelength of 1.55~\textmu m. First, we use the \texttt{lightbeam} \citep{lin21} Python package to simulate beam propagation through the PL in reverse and to determine the focal plane principal modes. Then we use the \texttt{HCIPy} (High Contrast Imaging for Python) package \citep{por18} to backpropagate focal plane principal modes to the pupil plane. Here we use an unobstructed circular aperture with a $D=$ 10~m diameter. The focal length is chosen to maximize the total coupling of an on-axis point source to all the supported LP modes, which gave a focal ratio of about 4.3. Note that the PLPMs are expected to be slowly varying functions of wavelength. Examining the effects of chromaticity is out of scope of this paper.

Using the numerically calculated pupil functions and their cross-correlations (Figure \ref{fig:ccpupil}), we derive complex visibilities ($\mathcal{V}_{ij}$) using Equation \ref{eq:vis_full} for a given mutual intensity at the pupil plane, $J_{\rm p}$. We then derive 15 squared visibilities ($|\mathcal{V}_{ij}|^2$) and 20 closure phases (argument of the bispectrum, $\phi_{ijk} = {\rm arg}(\mathcal{V}_{ij} \mathcal{V}_{jk} \mathcal{V}_{ik}^*)$). 

\subsubsection{PL visibilities for simple models}
\label{ssec:simplemodel}

We calculate PL visibilities for simple models: a uniform circular disk model and a binary model. For comparison with conventional separated-aperture interferometry, we consider placing pinholes at the locations where PL pupil mode amplitudes are at maximum: one pinhole at the center and five pinholes at the distance of $r_b = 3.4~$m from the center, forming a pentagon. 
All baselines are non-redundant in this configuration. 

\begin{figure}[t!]
    \centering
    \includegraphics[scale=0.47]{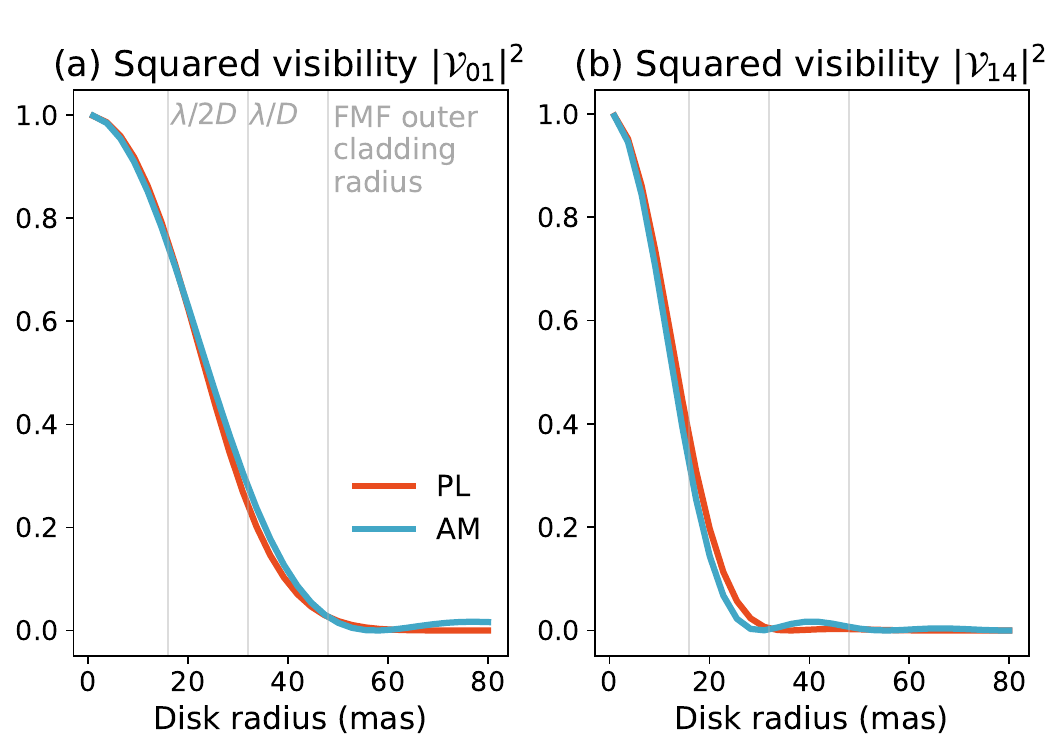}
    \caption{
    Simulated squared visibilities for PLs (red) and conventional interferometry (blue) for uniform circular disk model, as a function of disk radius. The squared visibilities of formed by two pairs are shown: mode 0 and 1, and mode 1 and 4. The telescope diameter $D = 10~$m and the wavelength $\lambda = 1.55$\textmu m are assumed.
    }
    \label{fig:models_disk}
\end{figure}

Figure \ref{fig:models_disk} shows expected squared visibilities for a resolved star (circular disk) model with uniform brightness distribution. Among the 15 squared visibilities, we display squared visibilities measured from subapertures 0 and 1 ($|\mathcal{V}_{04}|^2$) and 1 and 4, $|\mathcal{V}_{14}|^2$, as a function of circular disk radius. As the disk radius increases, squared visibilities deviate further from unity.  
The behavior of squared visibilities for the PL (red) is similar to that of classical interferometry (blue) at small angular size regime ($<\lambda/2D$), implying that PL visibilities have similar first-order responses to model parameters compared to classical visibilities.  

\begin{figure*}[hbt!]
    \centering
    \includegraphics[scale=0.42]{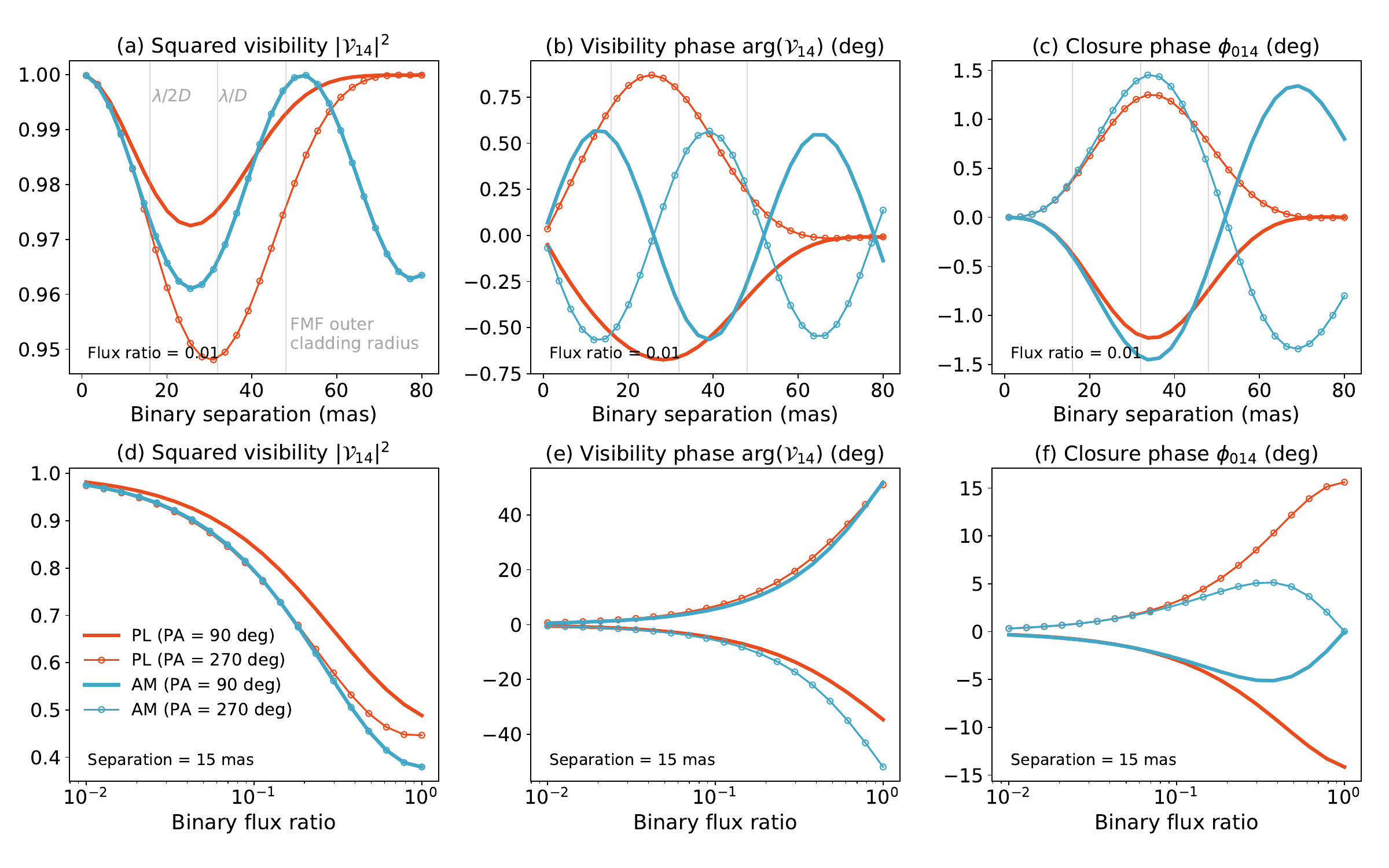}
    \caption{
    Simulated interferometric observables for PLs (red) and conventional interferometry (blue) for binary models, as a function of separation (top) and flux ratio (bottom). The cases for two position angles are shown, 90 deg (thick solid lines) and 270 deg (thin solid lines with circular markers). (a, d) Squared visibilities from modes 1 and 4. (b, e) Visibility phases from modes 1 and 4. (c, f) Closure phases formed from modes 0, 1, and 4. The telescope diameter $D = 10~$m and the wavelength $\lambda = 1.55$\textmu m are assumed.
    }
    \label{fig:models_binary}
\end{figure*}

Figure \ref{fig:models_binary} depicts the interferometric observables for a binary model, of which the primary star is placed at the center and the companion is placed off-center. The left and middle panels show squared visibilities and visibility phases measured from subapertures 1 and 4, and the right panels show the closure phases for the triangle formed by modes (0, 1, 4): argument of $\mathcal{V}_{01} \mathcal{V}_{14} \mathcal{V}^*_{04}$.

The upper panels show the observables calculated as a function of binary separation, for fixed flux ratio of 0.01, and with position angle of 90 degrees (thick lines) and 270 degrees (thin lines with circular markers).
In panels (a) and (b), the classical squared visibilities and visibility phases show sinusoidal patterns as a function of separation and the values for the two position angles are equal to each other. However, the PL squared visibilities and visibility phases of the two position angles are different, the case for 90 degrees position angle having larger signals than for the 270 degress position angle. This is because the cross-correlated PL pupil functions are extended and asymmetric, which is a deviation from normal AMI. The mutual intensity functions $J_p$ at ($u$, $v$) coordinates other than the baseline defined by the blobs contribute to the PL visibilities and helps break position angle degeneracy. Also, as the separation increases, the squared visibilities converge to unity and visibility phases converge to zero unlike for pinhole aperture masks, indicating the limited field of view. Also in panel (c), the closure phases of PL visibilities follow similar trends to classical visibilities but flatten and go to zero as separation increases beyond the size of the FMF entrance. The general behaviors are similar for other baselines and are shown in Figure \ref{fig:binary_all} (Appendix \ref{app:binary_all}).

The lower panels show the observables calculated as a function of binary flux ratio, for fixed separation of 15 mas. As binary flux ratio increases, the squared visibility and visibility phase signals increase, and the behavior of the PL visibilities is similar to classical visibilities. In contrast, classical closure phase signal goes to zero as binary flux ratio approaches unity. This is because classical closure phases can only sense asymmetries in the intensity distribution. That is, they probe non-linearity in the phase of the mutual intensity function (argument of Figure \ref{fig:plvis}(b)). When the binary flux ratio approaches unity, the phase of the mutual intensity function approximates to a sawtooth function, flipping its sign at $\lambda/2s$ 
where $s$ is the projected binary angular separation. Unless the projected binary separation is greater than $\lambda/2B$ where $B$ is the baseline length, closure phases for equal binary are zero because the sawtooth function is linear within a period. PL closure phase signals do not drop but steadily increase, as flux ratio approaches unity. 
This is because PL visibility phases are defined as the mutual intensity function weighted by the PL cross-correlated pupil functions, which are different for each baseline. The bispectrum phases do not sum up to zero even if the argument of the mutual intensity function is linear.
This implies that PL closure phases may be more sensitive to certain structures with symmetries, or at low contrast regime.

In summary, for disk and binary models, the PL interferometric observables behave similarly to the conventional interferometric observables with baselines defined by the distances between the peaks in the PL pupil functions, at scales $<\lambda/2D$. There are several differences that result from unique PL effective apertures. The PL visibilities are insensitive to scales larger than the size of the FMF entrance, which is roughly $1.5\lambda/D$ in our simulation. The change in squared visibilities under 180 degrees of rotation implies that PLs can break phase angle degeneracy. Moreover, PL closure phases have nonzero signals for symmetric structures. PL visibilities may be more sensitive to symmetric sources than conventional visibilities.

\section{Effects of Wavefront errors on measurement of PL visibilities}\label{sec:WFE}

In a realistic observation, the pupil phase is corrupted by time-varying WFEs.
For conventional AMI, where interferometric fringes are formed at the image plane, phase errors within a subaperture introduce redundancy errors \citep{rea88}. The fringes formed in the image plane correspond to the sum of the fringes formed by multiple redundant baselines that can be drawn within the subapertures. Thus, WFEs within a subaperture results in combining fringes incoherently. 
Eliminating the redundancy noise would in theory require infintesimally small subaperture sizes. Practically, the size of the subapertures is chosen to balance between throughput and redundancy effects as well as the Fourier coverage. For example, non-redundant aperture masks used with 10-m class telescopes typically have subaperture diameters smaller than 1\,m with number of holes $<$ 10 \citep{tut10b}. 

When the light transmitted by each subaperture feeds an SMF before interference, the subaperture redundancy errors can be reduced significantly. SMFs spatially filter the WFEs, rejecting aberrations other than the differential piston between the subapertures but at the cost of reduced and fluctuating coupling efficiency \citep{cou97, cou98, cha98, per03, per06, tut10, jov12, ber12}. This enables measurement of better fringe contrast and more stable closure phases. SMFs have been successfully used for combining beams from multiple telescopes \citep{cou97, per98, gra17} or combination with aperture masking techniques on a single telescope, called pupil remapping \citep{per06, tut10, jov12, vie23}.
Spatial filtering also increases coherence timescales, by averaging out temporal fluctuations within the subaperture. The atmospheric decorrelation
time increases with increasing the subaperture size when the pupil phases are averaged by spatial filtering \citep{con95, kel07,eis23}. This enables longer exposure times and thus observation of fainter targets.

PL interferometry leverages the spatial filtering nature of the SMFs since the PL converts few-moded telescope light into single-moded beams. However, there are a few differences with the techniques that inject telescope light directly into SMFs.
First, the pupil plane field is filtered by PL pupil functions instead of SMF pupil functions. The PL pupil functions have their unique amplitude and phase structures, while SMFs have a flat phase front Gaussian-shaped pupil function. Second, in the context of single-telescope interferometry, the telescope pupil is not divided into subapertures (pupil remapping or fibered AMI technique) but the entire telescope pupil is used. This allows better throughput compared to AMI, but significant redundancy in PL effective apertures is inevitable. In this section, we examine the effects of WFEs on PL visibility measurements by numerical simulations and compare them to traditional and fibered AMI techniques.

\begin{figure*}[hbt!]
    \centering
    \includegraphics[width=1\linewidth]{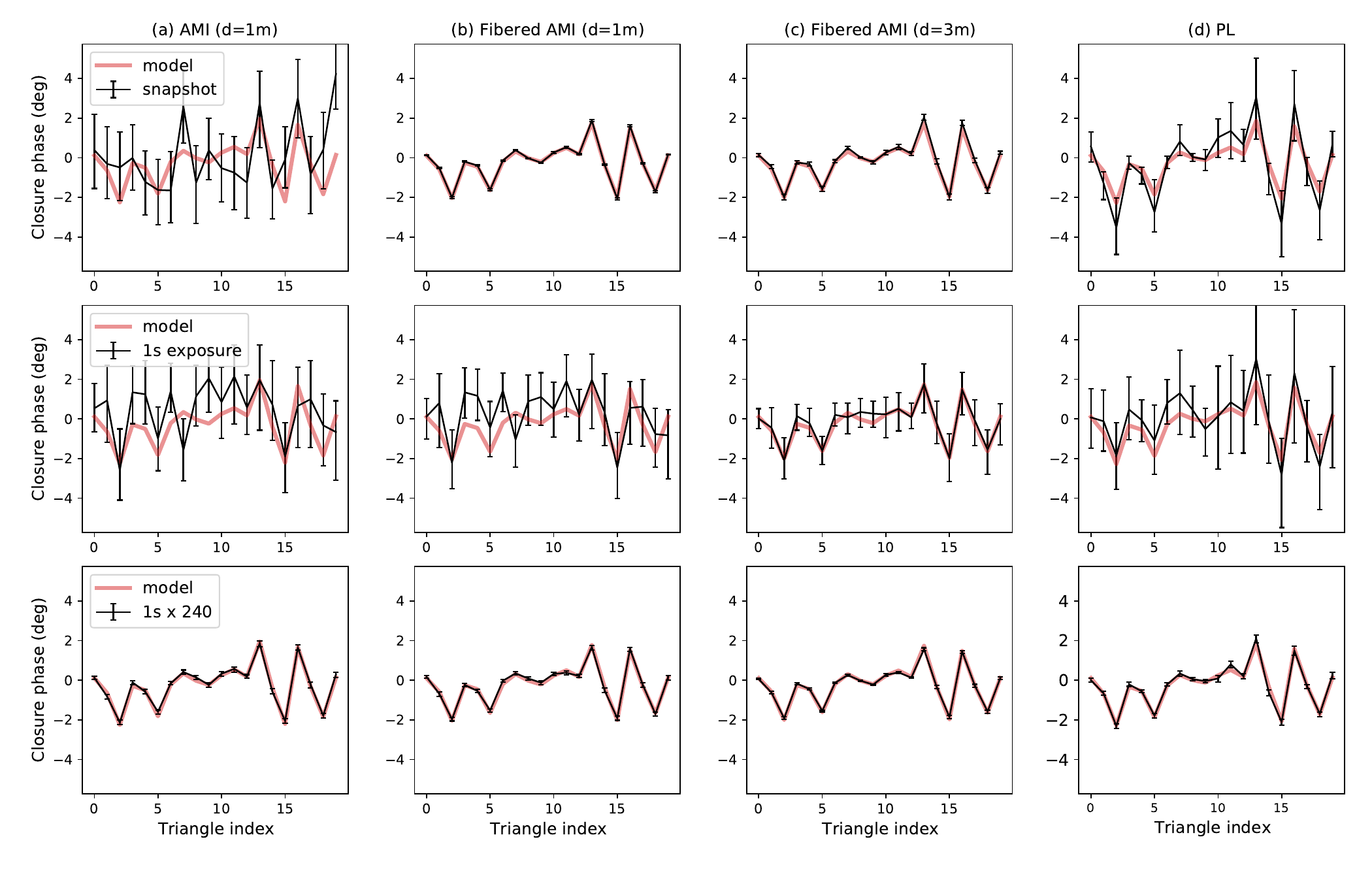}
    \caption{
    Simulated closure phase errors in presence of AO-residual WFEs (Strehl ratio = 0.54) observing binary stars with separation of 14 mas, flux ratio of 0.09, and position angle of 60 degrees, comparing four cases: (a) traditional AMI with 1\,m diameter subapertures, (b) fibered AMI with 1\,m diameter subapertures, (c) fibered AMI with 3\,m diameter subapertures, and (d) PL interferometry. Red lines show the model closure phases. (Top) Instantaneous closure phases and the error ranges are shown as black lines. Redundancy errors arise due to extended apertures. Spatial filtering effectively reduces redundancy errors. (Middle) Closure phases extracted from mock observations with 1 second integration time. The stability of closure phases is related to atmospheric coherence time, which increases with the size of the subapertures if the wavefront is spatially filtered. (Bottom) Averaged closure phases over 240 frames. Overall, PL closure phases have large redundancy due to the extended effective apertures but are relatively immune to effects of time-varying WFEs over an exposure, resulting in errors comparable to existing AMI techniques with subaperture diameter of 1\,m.}
    \label{fig:WFE}
\end{figure*}

\subsection{Simulated observations}\label{ssec:simobs}

We assume $D=$10~m unobstructed circular telescope pupil with adaptive optics. To simulate the effects of WFE, we generate Fourier-based Kolmogorov single-layer frozen-flow turbulent phase screens with 10\,m/s wind speed. The coherence length of the turbulence is set to 25\,cm. The aberrated wavefronts are partially corrected using a simulated closed-loop adaptive optics system implemented in \texttt{HCIPy} \citep{por18}. We assume using a deformable mirror with $30\times 30$ actuators across the pupil. The partially corrected phase maps are dominated by low-order aberrations and give average Strehl ratio of 0.54. We sample 30 AO-corrected phase maps over a period of 1 second (our assumed exposure time per frame), and generate 240 such independent realizations. Therefore our simulated phase screens consist of 240 independent exposures of the same atmospheric parameters, with each exposure consisting of 30 phase screens over one second.

We simulate binary star observations with three single telescope interferometry techniques, 1) PL interferometry as described in this study, 2) traditional AMI, and 3) fibered AMI. All the simulations are monochromatic with $\lambda = 1.55$\textmu m. 

\paragraph{PL interferometry}
For PL interferometry, the ABCD pairwise beam combination scheme is used. First, the mutual intensities are calculated by equation \ref{eq:J12} given source mutual intensity in the pupil plane and the PL pupil functions. We consider a standard 6-port PL as described in Section \ref{ssec:simulation}. The slow variation of PL pupil functions with wavelength is neglected in this study, which may be calibrated using the dispersed light in practice. Thus we assume the PL pupil functions we calculated for $\lambda = 1.55$ \textmu m (Figure \ref{fig:ccpupil}(a)). Effects of WFEs are applied to the PL pupil functions in equation \ref{eq:J12}. The intensities in ABCD outputs are calculated by equation \ref{eq:ABCD}. Then the complex visibilities are derived by equation \ref{eq:ABCD2vis}.

\paragraph{Traditional AMI}
We assume six hole aperture mask, with $d=1$\,m diameter circular holes positioned at the location where PL pupil mode amplitudes are at their maximum (Figure \ref{fig:ccpupil}(c)). Note that in practice the design of the mask (the number of the holes and their locations) could be optimized, but we use the mask configuration analogous to the PL pupil functions. For 10\,m telescope and 1\,m subaperture sizes, a few more subapertures could be placed for better Fourier coverage and throughput. Increasing the size of the suabpertures is unlikely because of the increased redundancy errors.
We use \texttt{HCIPy} to compute two dimensional Fourier transforms, from the pupil plane (panel (e) of Figure \ref{fig:ccpupil}) to the image plane, and to the Fourier plane (panel (f)).  To simulate binary observation, two planar wavefronts are generated and the corresponding fringe intensities are simply summed. The complex visibilities are extracted from the Fourier plane.

\paragraph{Fibered AMI}
To simulate fibered AMI, we use the same simulation scheme as the PL interferometry described above. The only difference is the effective pupil function. The telescope pupil is divided into 6 subapertures as in traditional AMI, then Gaussian SMF mode in the pupil plane is overlaid, with the focal length optimized to maximize the coupling efficiency. Existing fibered AMI instruments use subaperture sizes no larger than $d=1$m but with more numerous subapertures for better Fourier coverage. Given the number of subapertures and their configurations set by the PL pupil functions, thanks to spatial filtering, it is possible to increase the subaperture sizes up to $d=3$m within the telescope diameter of $D=10$m without overlapping, if not considering telescope obstructions. Thus, we consider two different subaperture diameters, $d=1$m and $d=3$m.

In the following subsections, we study the errors in closure phases in presence of static WFE (section \ref{ssec:snapshot}) and time-varying WFE (section \ref{ssec:exposure}), measured by the three techniques. For the binary parameters, we use separation of 14 mas, flux ratio of 0.09, and position angle of 60 deg. In calculating closure phase errors, we do not consider correlation of closure phase errors in this study. 

\subsection{Effects of redundancy}\label{ssec:snapshot}

If the integration time is shorter than the atmospheric decorrelation time, the fringes remain temporally correlated. The errors in the instantaneous closure phases are dominated by redundancy errors due to the finite size of the subapertures. The top panels of Figure \ref{fig:WFE} show simulated closure phases with the pupil plane phase aberrated using one of the simulated phase screens. The red solid line shows model closure phases. The error bars represent the 1$\sigma$ range of the closure phases generated from 240 independent phase screens. Comparing the (a) traditional AMI and (b) fibered AMI cases using identical subaperture sizes ($d=1$m), the advantage of spatial filtering becomes apparent. With SMF filtering, closure phase errors are significantly reduced. The errors increase if the subaperture sizes are increased ($d=3$m, (c)) even with spatial filtering. This is because the subaperture scale redundancy increases.

\begin{figure}[bt!]
    \centering
    \includegraphics[width=1.0\linewidth]{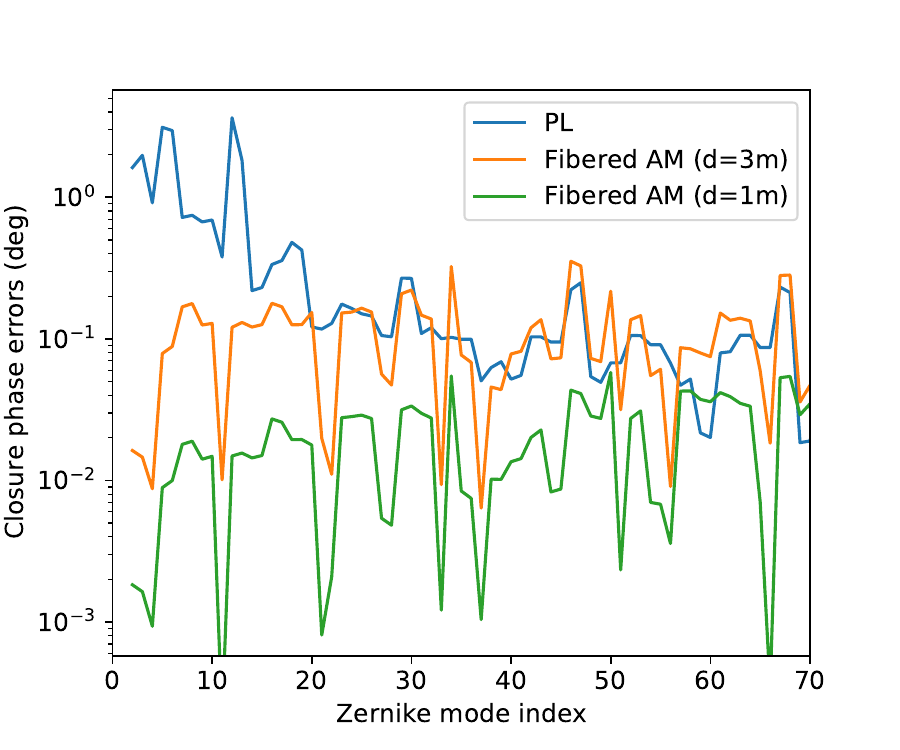}
    \caption{
    The closure phase errors in presence of random 0.5 radians rms jitter in each Zernike aberration mode. The errors averaged over all closure triangles are shown. Larger subapertures are more susceptible to redundancy errors. PLs have pronounced errors for low-order aberrations.}
    \label{fig:zernike}
\end{figure}

In (d), the instantaneous closure phases measured by PL are displayed. The error bars are much larger compared to the fibered AMI cases. This comes from the properties of PL effective apertures: extended effective apertures contribute to greater redundancy and the spatially low-order phase and amplitude structures make closure phases susceptible to low-order WFEs. Figure \ref{fig:zernike} shows the rms closure phase errors in presence of 0.5 radians rms error in each Zernike aberration modes. The errors are averaged over all 20 closure phases. We use the same input binary separation and contrast as in Figure \ref{fig:WFE}. The closure phase errors of fibered AMI are relatively flat across the Zernike modes. The $d=3$m case has about factor of 9 larger errors than the $d=1$m. PL interferometry shows larger closure phase errors in low-order Zernike modes regime ($<$20). This is because PL effective apertures are sensitive to low-order aberrations. Instead of spatially filtering the input wavefront with a flat wavefront, PLs filter the wavefront with unique phase structures, coupling low-order WFEs. We defer the exploration of practical calibration methods to future work. For high-order aberrations, PL closure phase errors have comparable errors with the $d=3$m fibered AMI case. Comparing the PL interferometry with traditional AMI technique, the errors are still smaller due to spatial filtering.

\subsection{WFE effects over an exposure}\label{ssec:exposure}

We next consider stacking fringes (for traditional AMI) or ABCD outputs (for fibered AMI and PL interferometry) over a 1-second integration time. The middle panels of Figure \ref{fig:WFE} show the simulated closure phases extracted from one of the 1-second mock observations. The error bars are calculated from 1$\sigma$ range of the 240 realizations of 1-second mock observations. Incoherently summing the fringes/outputs, closure phases are corrupted. The fibered AMI with $d=3$m case has smaller errors than the $d=1$m case, which can be attributed to enhanced atmospheric decorrelation time ($\sim d/v$, where $v$ is the wind speed). Larger subaperture sizes help in the long-exposure regime, with spatial filtering. The closure phases of the PL interferometry case (d) are of similar order compared to the fibered AMI with $d=1$m (b).

In the bottom panels of Figure \ref{fig:WFE} illustrate the closure phases averaged over 240 frames each consisting of 1 second integration. The error bars are the estimated 1$\sigma$ range of the mean. Averaging closure phases over multiple frames, the closure phases can be accurately recovered. The error range of the PL interferometry is comparable to that of the fibered AMI with $d=1$m.

In summary, PL interferometry suffers from large redundancy errors due to the extended aperture and sensitivity to low-order aberrations, but is relatively stable in the long exposure regime due to the spatial filtering and the extended effective aperture. Therefore, PL interferometry may provide better efficiency for observing faint targets where sufficient integration time is needed. Practical calibration strategies for low-order aberrations and consideration of closure phase covariances are left for future work. Moreover, PLs are expected to have better throughput than AMI techniques because the entire telescope pupil is used. In the following section, we perform simulated observations including photon noise and characterize the regime in which PL interferometry is powerful.

\section{Potential of PL interferometry}\label{sec:potential}

In the previous sections, we have defined PL visibilities, compared them to conventional visibilities, and studied the sensitivity of closure phases to WFEs. In this section, we explore the potential of PL interferometry using all the observables (relative intensities, squared visibilities, and closure phases) and introducing photon noise. We describe the methodology of generating mock data in \S\ref{ssec:mockdat}. Then we investigate the potential of PL interferometry using the mock data for two approaches: model fitting (\S\ref{ssec:param}) and nonparametric image reconstruction (\S\ref{ssec:nonparam}). 

\subsection{Mock data generation}\label{ssec:mockdat}

We generate mock data for three cases, PL interferometry, fibered AMI with $d=1$m, and fibered AMI with $d=3$m as described in \S\ref{ssec:simobs}. To simulate wavefront errors, as discussed in the aforementioned section, we use the 240 1-second AO-corrected phase screens that give average Strehl ratio of 0.54. In a real observation, the observed quantities need to be calibrated against an observation of an unresolved source. Therefore, we assign 120 exposures to the science target observation and the other 120 exposures to the calibrator star observation. 
The visibility and intensity information are extracted from each ABCD frame and then are converted to following observables: relative intensities, squared visibilities, and closure phases. The extracted observables are averaged over 120 exposures both for the science target and the calibrator star.

\paragraph{Relative intensities}
For PL interferometry, relative intensities depend on the input scene which can be used to constrain the input scene in addition to visibility information. We calibrate the relative intensities simply by subtracting by the difference between the observed relative intensities of the calibrator star and the relative intensities for an unaberrated point source \citep{kim22}. This eliminates biases introduced by jitter effects independent of the input scene, but the systematic error term that depend on the input scene remains uncalibrated. Practical calibration methods are left for future work. The errors in relative intensity estimates are dominated by the calibration error, originating from using different phase screens for the science target and the calibrator star.

\paragraph{Squared visibilities}
Dividing the observed squared visibilities by those of the calibrator star's, accounting for the instrument transfer function, gives unbiased estimate of the squared visibilities. The squared visibility errors are also dominated by calibration errors.

\paragraph{Closure phases}
Closure phases need not be calibrated because we did not introduce instrumental closure phase errors nor quasi-static WFEs in the simulation. Thus, we do not attempt to calibrate closure phases against those of the calibrator star's.  The closure phase errors are dominated by incoherent ABCD output intensity integration over an exposure (\S\ref{ssec:exposure}).

\begin{figure*}
    \centering
    \includegraphics[width=1\linewidth]{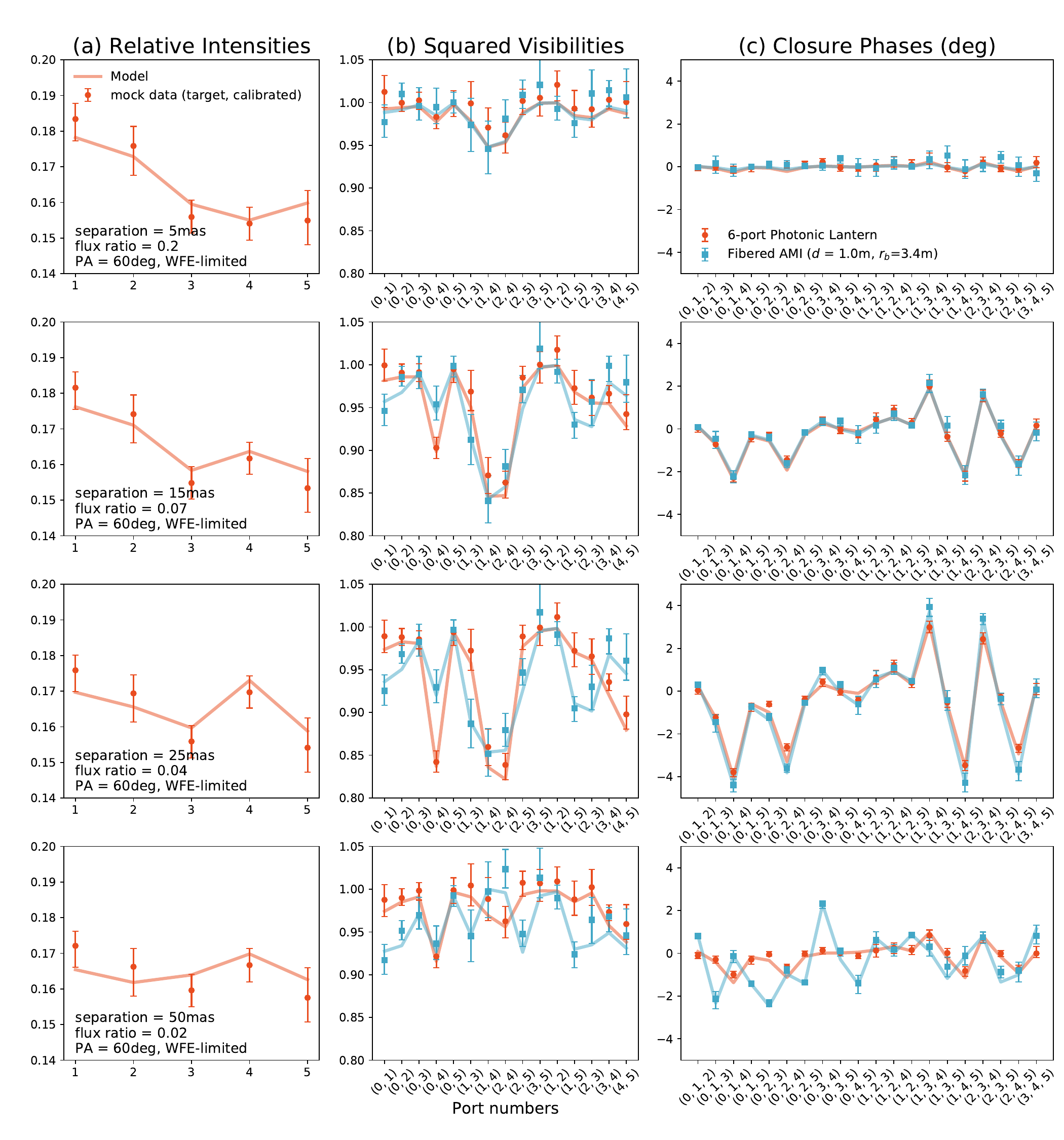}
    \caption{
    Simulated observables for binary point sources calibrated against mock observation of an unresolved star, which include (a) relative intensities (total intensity normalized to 1), (b) squared visibilities, and (c) closure phases. 
    The mock observation consists of 120 1-second exposures and no photon noise is added, assuming the WFE-limited case. Red circles show the case for PL interferometry and blue squares show the case for the fibered AMI ($d=$1m) analog. The models are shown as solid lines.}
    \label{fig:mock_data}
\end{figure*}

In the WFE-limited regime, the squared visibility and closure phase errors of the PL interferometry and the fibered AMI with $d=1$m are comparable. Fibered AMI with $d=3$m has slightly smaller errors than the $d=1$m case (see \S\ref{sec:WFE}). Figure \ref{fig:mock_data} shows example mock observations of binary point sources with four different binary separations and contrasts, with no photon noise added. The red circles and blue squares display the mock data for PL interferometry and fibered AMI ($d=1$m case), respectively. See \S\ref{ssec:param} for more details.  

To simulate photon noise-limited case, we add photon noise in each ABCD frame. 
The throughput of PL and fibered AMI depends mainly on the coupling efficiency, the ratio of the light that gets coupled into the fiber to the light incident on the telescope pupil, which correlates with the Strehl ratio \citep{jov17, lin22a, lin22spie}. For the simulated phase screens,
we compute the average coupling efficiency of the 6-port PL as 54\% and the fibered AMI as 3\% and 27\% for $d=1$m and $d=3$m, respectively. 
Once the light is coupled into the fiber entrance, the transition losses are assumed to be negligible. We do not account for the throughput of the integrated optics beam combiner.

\subsection{Parametric Modeling --- A case for the Binary Model}\label{ssec:param}

Once observations are complete and visibilities are computed, a straightforward way to interpret the signals is to fit a model to the observed quantities and evaluate the significance. Models of interferometric observables can be constructed from a model mutual intensity function $J_{\rm p}(u,v)$ and information on pupil functions (Equation \ref{eq:vis_full}). 
In this subsection, we show an example for binary models. 
The parameters of binary models include separation, flux ratio, and position angle. 
We compute significances and posterior distributions for a range of binary separations and flux ratios and 
discuss how PL interferometry observables are useful for constraining the parameters.

Figure \ref{fig:mock_data} displays mock data of binary systems in WFE-limited regime. Four binary separation regimes are explored, 5 mas (0.16\,$\lambda/D$), 15 mas (0.47\,$\lambda/D$), 25 mas (0.78\,$\lambda/D$), and 50 mas (1.56\,$\lambda/D$), with flux ratios of 0.2, 0.07, 0.04, and 0.02, respectively. The separation and flux ratio parameters are chosen to have (separation $\times$ flux ratio) constant. 
For the small separation cases (5\,mas and 15\,mas), the closure phases for the PLs and for the fibered AMI ($d=1$m) are similar, as suggested from Figure \ref{fig:models_binary}. 
As increasing the separation, PL closure phase signals become smaller. 
The relative intensity signals may provide more constraints on the source at very small separations (e.g., 5 mas case) because the relative intensity signals depend linearly on binary separation in this regime \citep{kim22}, where visibility signals have nearly quadratic dependence (Figure \ref{fig:models_binary}).

\begin{figure*}[hbt!]
    \centering
    \includegraphics[width=1\linewidth]{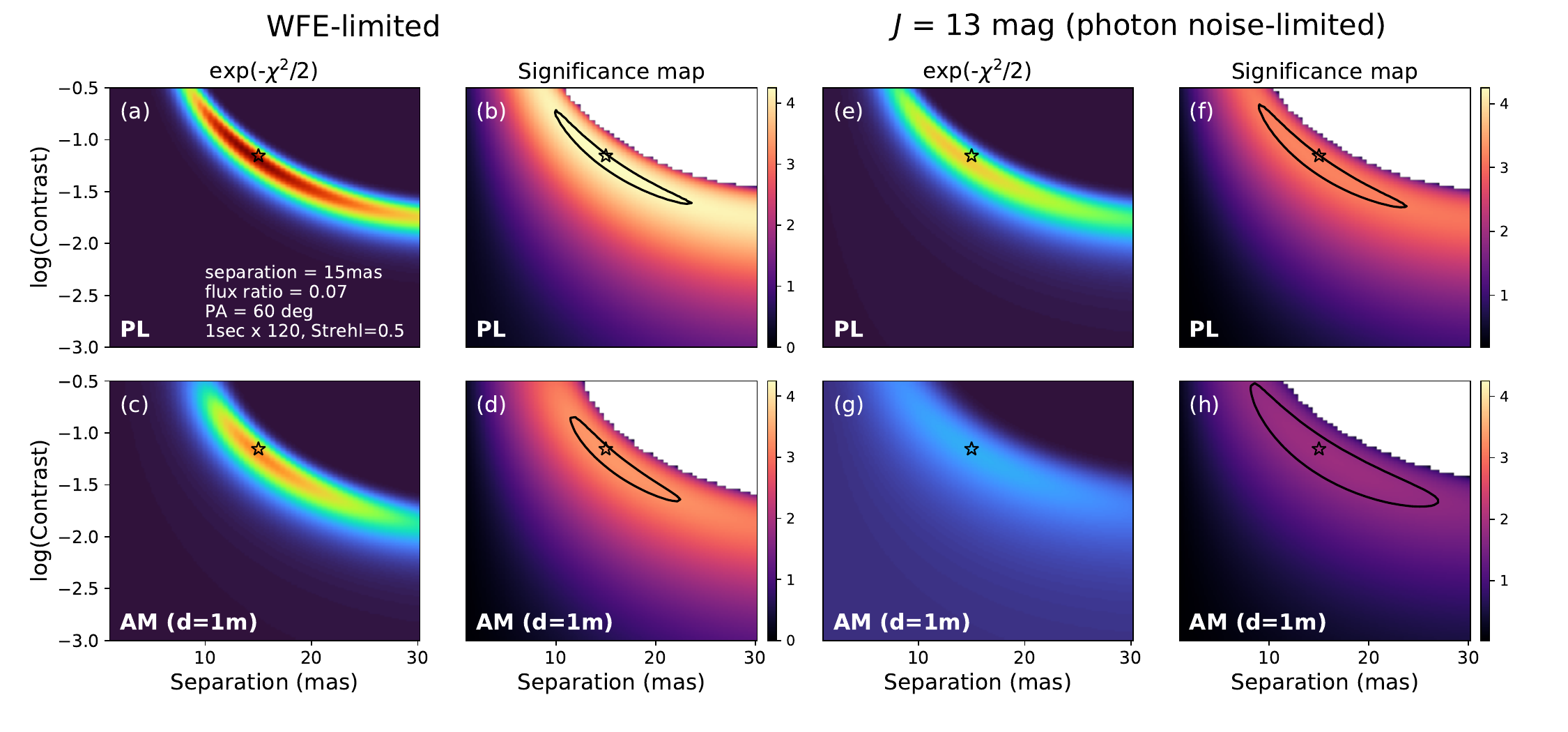}
    \caption{Posterior distributions ((a), (c), (e), (g)) on separation and contrast and significance maps ((b), (d), (f), (h)) of the mock observation shown in the second row of Figure \ref{fig:mock_data}. The top and bottom rows display those calculated for PLs and aperture masks, respectively. The left panels show the WFE-limited regime, and the right panels show the photon noise-limited regime, observing 13th-magnitude binaries. The position angle is fixed to the true value, 60 deg. The star symbol shows the true binary parameters. The plots are on the same color scale. }
    \label{fig:significance_maps_a}
\end{figure*}

\begin{figure*}[hbt!]
    \centering
    \includegraphics[width=1\linewidth]{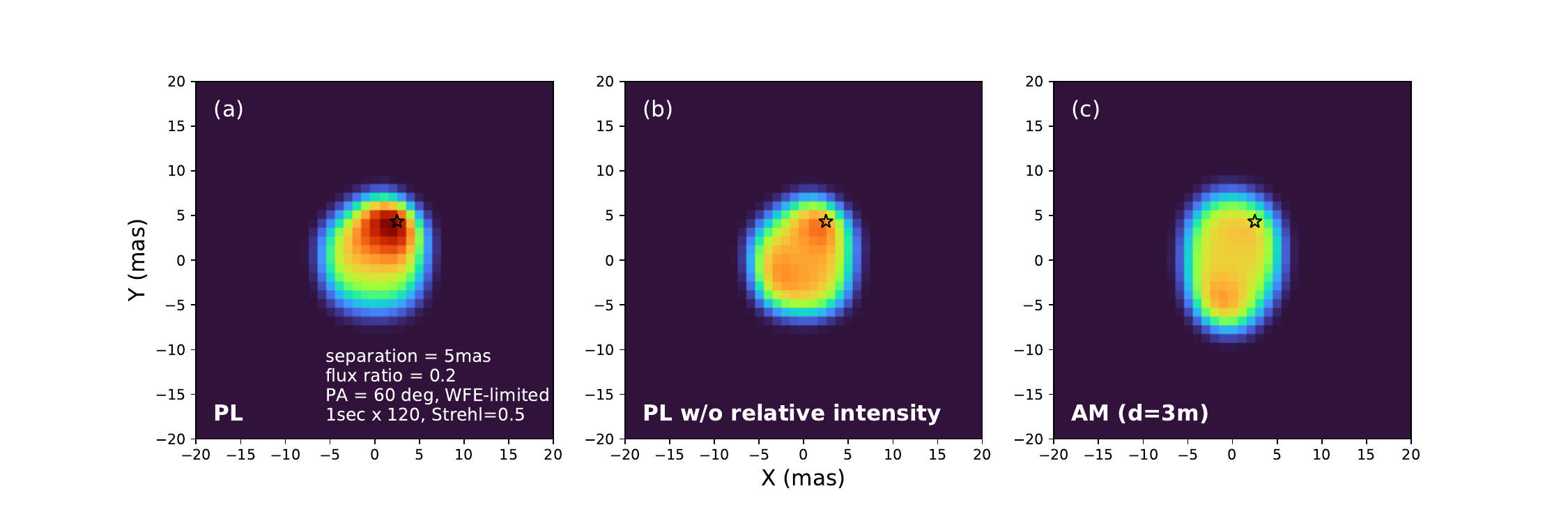}
    \caption{Posterior distributions of the companion location in a WFE-limited mock observation of binaries with 5\,mas separation, the first row of Figure \ref{fig:mock_data}. The posterior distributions are calculated using (left) all PL observables, (middle) PL observables without relative intensities, and (right) fibered AMI ($d=3$m). The contrast is fixed to the true value, 0.2. The star symbol shows the true companion location. In the small angular separation regime, the relative intensity information helps break the position angle degeneracy and localize the companion. The three plots are on the same color scale.} 
    \label{fig:significance_maps_c}
\end{figure*}

The panels (a) and (c) in Figure \ref{fig:significance_maps_a} display the estimated posterior distributions for the mock data shown in the second row of Figure \ref{fig:mock_data}: separation of 15 mas (0.47\,$\lambda/D$) and flux ratio of 0.07. The posterior distributions are calculated as $\propto \exp{(-\chi^2/2)}$ on a grid of separation and flux ratio. $\chi^2$ is the chi-squared, the sum of squared differences between the mock data and the true model, divided by the variance. We assume independent Gaussian uncertainties for the interferometric observables. The star symbol shows the true value for separation and flux ratio. The position angle is fixed to 60 deg in this map. There is a degeneracy between the separation and contrast for binary models with separations within the classical diffraction limit. 

The panels (b) and (d) in Figure \ref{fig:significance_maps_a} depict the significances calculated on the same grid, which are estimated as
\begin{equation}
    \sigma = \sqrt{\chi^2_{\rm null} - \chi^2}
\end{equation}
as in \citet{wil16} and \citet{sal19}. $\chi^2_{\rm null}$ represents the chi-squared assuming the true model as an unresolved point source. The black solid contour line corresponds to the level where $\sigma^2$ drops by 0.5.
We find a slightly higher level of significance for the PL than for the aperture mask in this case and the size of parameter space within the contour is similar.

In the right panels of Figure \ref{fig:significance_maps_a}, we repeat the same procedure but with photon noise, assuming a 13th-magnitude point source in the $J$-band (bandwidth of $\lambda/\Delta\lambda= 300$). The significance levels are lower than in the previous case. 
The posterior distributions and significances reflect the differences in throughput, indicating better performance for the PL than for the AMI. 

Figure \ref{fig:significance_maps_c} shows the posterior distributions of the companion location for the mock data shown in the first row of Figure \ref{fig:mock_data}: separation of 5 mas (0.16\,$\lambda/D$) and flux ratio of 0.2.
The contrast is fixed to the true value in this map, 0.2. The mock observations include (a) PL observables, (b) PL observables but without information on relative intensities, and (c) fibered AMI ($d=3$m) observables. 
In this small separation regime, the closure phase signals are very small, thus there is a significant degeneracy in the position angle. For PLs the distinct squared visibilities upon 180 degrees rotation help break the degeneracy but the squared visibility signals are also small.  
The information from relative intensities further helps break the degeneracy, resulting in a single peak in the posterior distribution.

\begin{figure}
    \centering
    \includegraphics[width=0.8\linewidth]{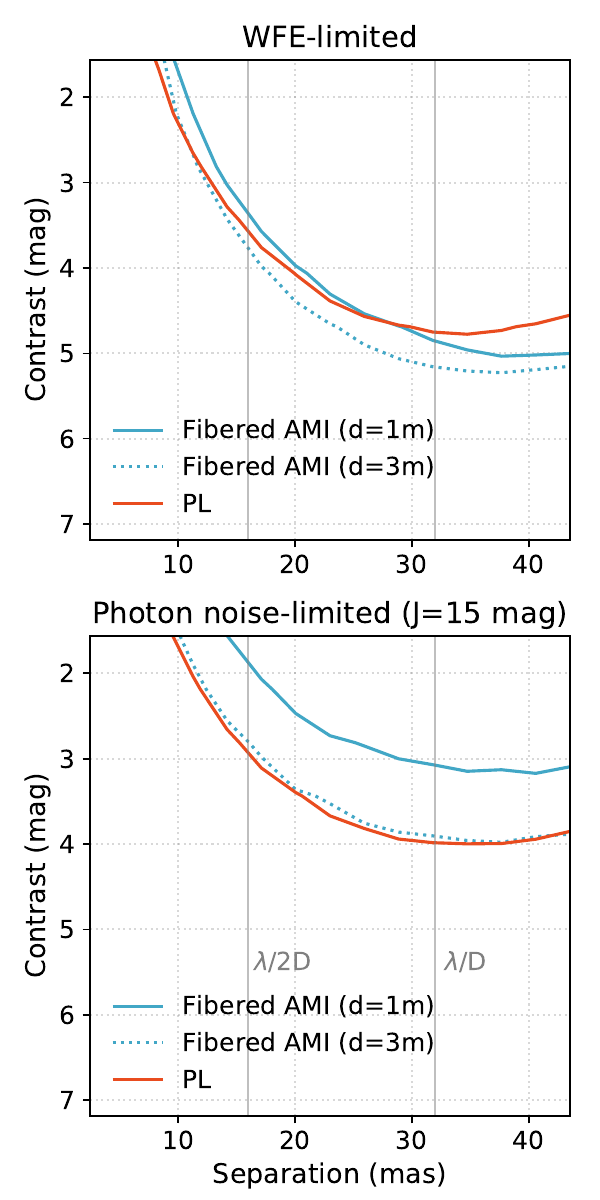}
    \caption{Contours showing significance level of 3$\sigma$ for companion detection, comparing PL interferometry and fibered AMI ($d=1$m and $d=3$m).
    (Top) The significance levels in the WFE-limited regime. (Bottom) The significance levels in the photon noise-limited regime, observing the source brightness of $J=$15 mag.
    In photon-noise limited regime, throughput plays a major role in determining the significance levels, resulting in distinct curves between PLs and aperture masks.}
    \label{fig:binary_significance}
\end{figure}

Figure \ref{fig:binary_significance} shows contours of constant significance level ($3\sigma$) calculated on a grid of binary separation and contrast, for both the PL and the fibered AMI with $d=1$m and $d=3$m. The significances are averaged over random values of position angles. 
The top and bottom panels show the WFE-limited case (assuming no photon noise) and the photon noise-limited case ($J=$\,15\,mag), respectively. 
In the WFE-limited regime, the fibered AMI with $d=3$m reaches a deeper contrast limit than the fibered AMI with $d=1$m due to the enhanced effective atmospheric coherence time (see \S\ref{ssec:exposure}). In the photon noise-limited regime, the differences are more pronounced, due to the differences in the throughput.
The significance level of the PL interferometry in the WFE-limited case is nearly in between the fibered AMI with $d=1$m and $d=3$m at separations smaller than $\lambda/D$. PL significance reaches a deeper contrast limit at very small separations, due to larger closure phase signals at small flux ratio regime and larger relative intensity signals at small separation regime.
At separations larger than $\lambda/D$, the contrast limit decreases for PLs, reflecting decreased sensitivity due to the physical dimension of the FMF entrance at the focal plane. PLs can reach greater contrast in the photon noise-limited regime due to the better throughput than fibered AMI techniques.

\subsection{Nonparametric Image Reconstruction}\label{ssec:nonparam}

Interferometric methods in the near-infrared enable spatially resolving the inner regions ($<$ few au) of circumstellar environments of nearest star forming regions \citep{wil16, gra19}. To test the potential of image reconstruction when using PL-measured visibilities, we run monochromatic image reconstruction simulations using the Markov Chain Monte-Carlo (MCMC) algorithm. 
MCMC with simulated annealing image reconstruction techniques have been successfully used in optical inferometry, by the Markov Chain Imager \citep[\texttt{MACIM};][]{ire06} and \texttt{SQUEEZE} \citep{bar10}.

\begin{figure*}[hbt!]
    \centering
    \includegraphics[scale=0.45]{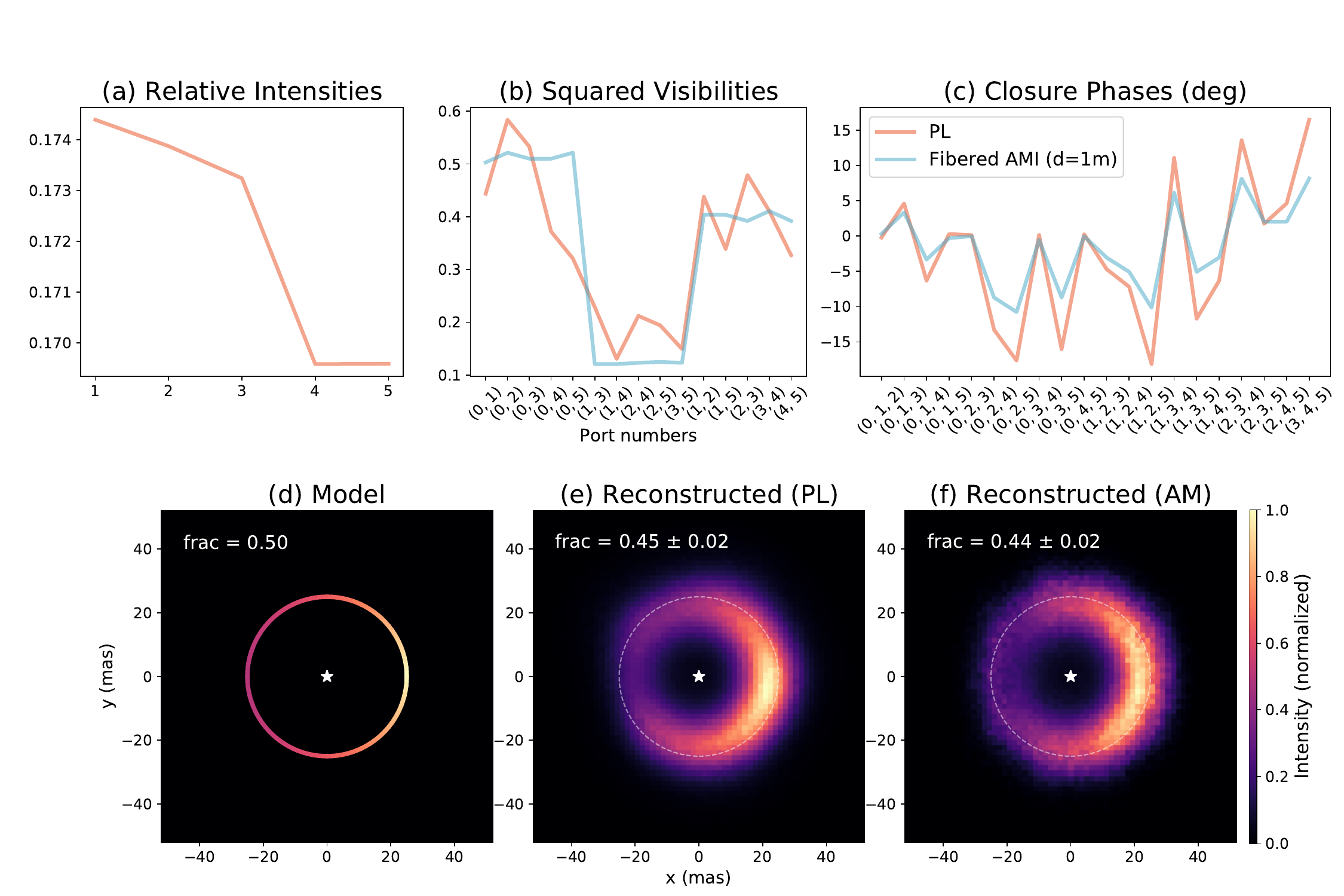}
    \caption{An example image reconstruction of circumstellar environment. The model, depicted in (d), comprises a central unresolved star contributing half of the total flux and an asymmetric circular ring-shaped disk with a radius of 25 mas. The top row displays the interferometric observables computed for this model using both the PL and fibered AMI ($d=1$m). The resulting images, obtained through the MCMC algorithm, are shown in (e) for PL interferometric observables and in (f) for fibered AMI. Both the PL and the AMI recover the asymmetric ring which is more extended than the interferometric resolution ($\lambda/2D$).}
    \label{fig:squeeze1}
\end{figure*}

\begin{figure*}[hbt!]
    \centering
    \includegraphics[scale=0.45]{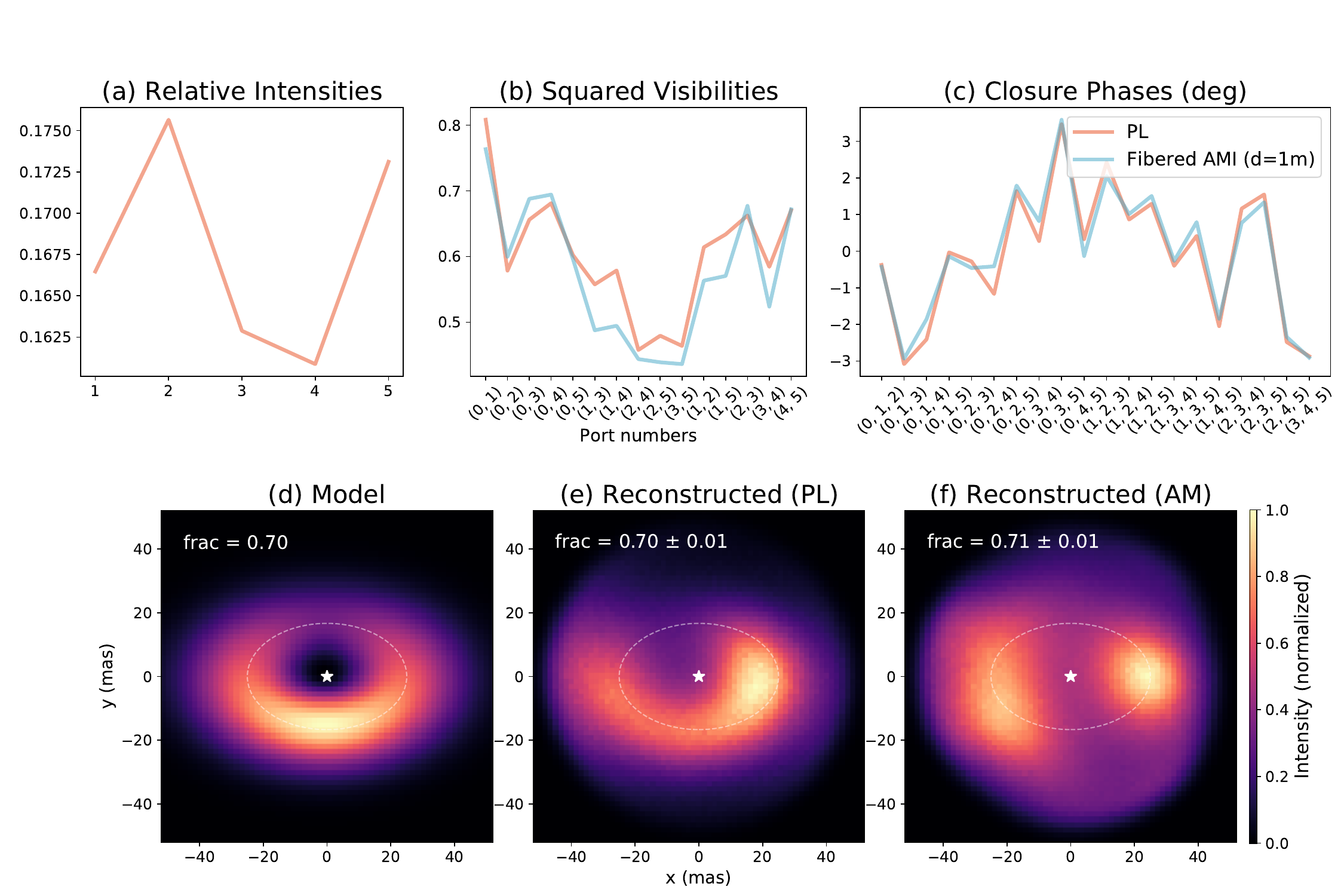}
    \caption{Same as in Figure \ref{fig:squeeze1} but with another model, a smooth asymmetric disk with a bright rim along the semi-minor axis. The bright rim is recovered in PL reconstructed images but not in AMI reconstructed images.}
    \label{fig:squeeze2}
\end{figure*}

Panel (d) in Figure \ref{fig:squeeze1} shows an example scene we aim to recover. The image includes an unresolved star at the center (indicated by star symbol) surrounded by a thin circumstellar disk. The model disk has a radius of 25 mas ($=0.78\lambda/D$), larger than the interferometric resolution limit, and a skewness of 0.5. The flux of the star constitutes 50\% of the total flux. The top panels present the interferometric observables without noise. 
Note that PL closure phases have larger signals than those of the fibered AMI. This is because of some degree of symmetry in the input scene. The asymmetric ring is qualitatively similar to a central source with collection of two companions positioned at equal distances along +x and -x. The skewness determines the relative flux between the companions. If the flux contrast between the central star and the ring is moderate, the situation becomes similar to binary stars with a moderate contrast case (\S \ref{ssec:simplemodel}). Therefore PL closure phases are greater than classical closure phases as can be inferred from Figure \ref{fig:models_binary}(f).
We generate mock data using the same turbulence phase screens as in \S \ref{ssec:mockdat} and the same calibration method. We do not include photon noise in this mock observation. For the fibered AMI, we use the $d=1$m case.

We then run MCMC with simulated annealing algorithm to recover the circumstellar disk. To effectively image the environment of the central unresolved star, a point source model is used as in \citet{klu14}. An $M \times M$ pixel image is described by a collection of elements, which may be pixel fluxes or the fraction of the central star's flux to the total flux. On each step, the flux elements move their location across the image grid or change central star's flux based on the Metropolis acceptance probability. We use $M=65$ and 1.6\,mas pixel scale. We adopt the temperature schedule used in \texttt{MACIM} and \texttt{SQUEEZE}.
We use a uniform circular prior with smooth edges that restricts the field of view to 47\,mas in radius. It is important to limit the field of view because PL visibilities are insensitive to the fluxes outside the PL entrance, which may result in lingering fluxes in the edges of reconstructed images. We do not include other regularizations than the positivity, which is inherently sufficient. The target reduced $\chi^2$ is set to 1 and the parameters of the temperature schedules are adjusted for convergence.

Panels (e) and (f) of Figure \ref{fig:squeeze1} display reconstructed images of environments of the central source, for PL and fibered AMI ($d=1$m), respectively. Both the PL and the AMI recovers the morphology of the circumstellar component which is extended beyond the interferometric resolution limit, although the fraction of the central star's flux is slightly underestimated. 

Figure \ref{fig:squeeze2} shows another example of image reconstruction. In this case, the disk has a smooth intensity distribution, with a semi-major axis of 25 mas and an axis ratio of 1:1.5 (panel (d)). The flux of the star constitutes 70\% of the total flux. The top panels show the interferometric observables. The signals are smaller than the former case in Figure \ref{fig:squeeze1} because of the decreased relative flux of the circumstellar component. Thus, the effects of WFEs are more prominent, as seen from distortions and the brightened edge in reconstructed images, panels (e) and (f). While the PL recovers the bright component along the semi-minor axis, the AMI only recovers the extended emission along the semi-major axis (beyond the interferometric resolution). 
The mutual intensity variation along the y-axis resembles that of a moderate contrast binary with a small separation, which classical closure phases are insensitive to. Further increasing the flux fraction of the central star to greater than 90\%, the mutual intensity function exhibits more nonlinear behavior (similar to higher contrast binary), resulting in sensitivity to structures of $<\lambda/2D$ of PLs and AMI becoming comparable.

\section{Discussion}\label{sec:discussion}

We presented a concept of using photonic lanterns for interferometric imaging in \S \ref{sec:concept}. Interpreting pupil plane lantern principal modes (pupil plane PLPMs) as effective apertures, the visibilities measured from interfering SMF outputs are mutual intensities weighted by cross-correlation of pupil functions, integrated over $u-v$ plane. We showed that the PL visibilities are similar to separated-aperture visibilities in angular scales $<\lambda/2D$. PL squared visibilities break phase angle degeneracy and closure phases have signals for symmetric structures due to  unique effective apertures of PLs. We examined the effects of WFEs on measurement of PL closure phases in \S \ref{sec:WFE}. PL closure phases are sensitive to low-order aberrations but are relatively immune to effects of time-varying WFEs over an exposure thanks to spatial filtering. 
In \S \ref{sec:potential}, we presented simulated observations and showed the potential of model fitting and non-parametric interferometric image reconstruction with PL visibilities. We found two main benefits of PL interferometry from simulated observations.
First, the high throughput of PLs makes them particularly efficient in the photon noise-limited regime (Figure \ref{fig:binary_significance}). PL interferometry would require a shorter integration time than aperture masking interferometry allowing efficient observation of faint targets and high resolution spectroscopy. 
Second, PL closure phases' sensitivity to symmetric scenes may be beneficial for resolving small scale structures ($\sim \lambda/2D$) when the flux ratio between the central star and the circumstellar component is moderate. This is the regime where PL interferometry can access scales that are too small for sufficient aperture masking interferometry constraints. Standard aperture masking interferometry can fill a gap between the regime PL interferometry is insensitive due to the limited field of view and where traditional imaging techniques have deep contrast.

Furthermore, the single-moded beams from the beam combiner outputs can be fed into a high-resolution spectrometer. 
This would enable high spectral resolution interferometric data and expand the detection limit to even smaller angular scales, using spectro-interferometry \citep{des07}. This may be cost-effective as it can be integrated with existing SMF-based spectrometers. Moreover, the spatial selectivity of the PL is a major advantage in the case of crowded fields as well as with reducing the background noise. 

While this study is based on a simulated standard 6-port PL, one can consider PL interferometry with various lantern designs. The PLPMs and thus the $r_b$ value are functions of the lantern geometry, such as core sizes, core spacings, and taper length \citep{lin23}. Design of the lantern may be optimized for desired Fourier coverage and better sensitivity at high spatial frequencies. In addition, a PL with a larger mode count may provide increased Fourier coverage, a larger field of view, and a higher throughput. However, there are several challenges to be taken into account. As light splits into a greater number of ports, the intensity in each port will decrease, increasing the photon noise in each output. In addition, if considering a pairwise beam combination scheme, the required number of ABCD beam combiners and outputs scales with the number of SMF outputs $N$ as $\propto N^2$, increasing complexity. 

There is much more design space one can explore. Potentially the PL may be simultaneously used for both science and focal-plane wavefront sensing \citep{nor20, lin22a}. By adjusting field amplitudes and phases, nulling interferometry of our beam combiner outputs may be performed to achieve a higher contrast, which is an addition to the concept of using a mode-selective PL for nulling on-axis starlight \citep{xin22, xin23} or using a hybrid PL with a vortex mask \citep{lin23}. Also, the input electric field may be modulated using a deformable mirror to decompose coherent light (starlight) and incoherent light (pairwise probing approach; \citet{giv11}) to achieve a higher contrast. If using an oversampled lantern of which the number of SMF outputs is larger than the number of supported modes, the SMF output intensities may contain interferometric information on the source and PL interferometry may be performed directly from SMF output intensities without a beam combiner. One can envision more advanced or more practical beam combiner designs, but a comprehensive study is outside the scope of this paper.

There are several complications to be addressed in future studies. 
The complex PL mode structures need to be accurately determined in the laboratory for precise model fitting and image reconstruction. 
The effects of obstruction in the telescope pupils need to be considered.
Additionally, the potential correlations in uncertainties in PL visibilities, caused by overlapping PL modes structures, should be examined to ensure accurate model fitting. Practical complications include accounting for the variation in throughput of different PL outputs, balancing the intensities in beam combination, and dealing with chromaticity and polarization.

\section*{Acknowledgements}
We thank the anonymous reviewer for their valuable feedback that helped improve the paper. We thank Jessica Lu for helpful suggestions. This work is supported by the National Science Foundation under Grant No. 2109231, 2109232, 2308360, and 2308361.

\vspace{5mm}

\software{astropy \citep{astropy:2013, astropy:2018, astropy:2022},  
          NumPy \citep{harris2020array},
          SciPy \citep{2020SciPy-NMeth},
          Matplotlib \citep{Hunter:2007},
          HCIPy \citep{por18},
          lightbeam \citep{lin21}}

\appendix

\section{PL visibilities for binary model}\label{app:binary_all}
\begin{figure}[hbt!]
    \centering
    \includegraphics[width=0.96\linewidth]{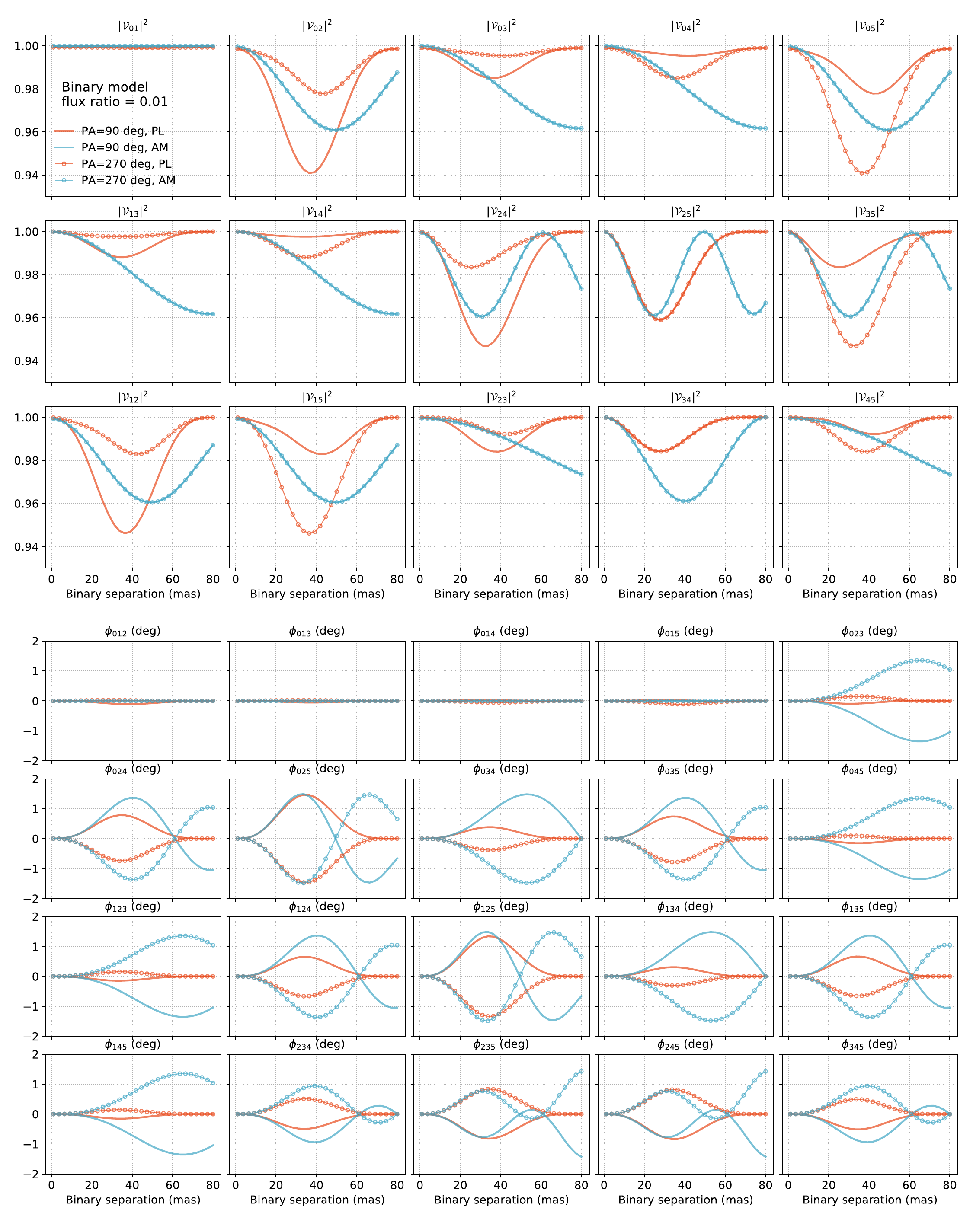}
    \caption{Simulated interferometric observables for PLs (red) and conventional interferometry (blue) for binary models, as a function of separation. The flux ratio is fixed to 0.01. The cases for two position angles are shown, 0 deg and 180 deg. All the 15 squared visibilities (top) and the 20 closure phases (bottom) are displayed.} 
    \label{fig:binary_all}
\end{figure}

All the simulated 15 squared visibilities (top) and 20 closure phases (bottom) for binary models of position angles 0 deg and 180 deg are shown in Figure \ref{fig:binary_all}. The general behaviors are similar with the Figure \ref{fig:models_binary}.

\bibliography{sample631}
\bibliographystyle{aasjournal}

\end{document}